\documentclass[referee]{aa} 
\usepackage{graphicx}
\usepackage{txfonts}
\newcommand{\hm}{H$_{2}$}

\newcommand{\dm}{D$_{2}$}

\begin{document}

\title{The role of carbon grains in the deuteration of \hm}  
\author{S. Cazaux\inst{1}, P. Caselli\inst{2}, V. Cobut\inst{3}, J. Le Bourlot\inst{4}} 
\offprints{cazaux@astro.rug.nl}
\institute{Kapteyn Astronomical Institute, PO box 800, 9700AV Groningen, The Netherlands\\
\and 
School of Physics and Astronomy, University of Leeds, LS2 9JT, Leeds, UK\\
\and
Laboratoire pour l'Etude du Rayonnement et de la Mati\`ere, UMR 8112 du CNRS, Observatoire de Paris et Universit\'e de Cergy Pontoise, 5 Mail Gay-Lussac, F-95031 Cergy-Pontoise Cedex, France\\
\and
LUTH, UMR 8102 CNRS, Universite Paris 7 and Observatoire de Paris, Place J. Janssen, 92195 Meudon, France} 

\date{Received 04 september 2007; accepted 20 february 2008}

\abstract {}{The production of molecular hydrogen and its deuterated
forms onto carbonaceous dust grains is investigated in detail. The
goal of this study is to estimate the importance of the chemistry
occuring on grain surfaces for the deuteration of \hm. Furthermore, we
aim to find a robust and general surface chemical model which can be
used in different astrophysical environments.} {Surface processes are
described for the cases of graphitic and amorphous--carbon grains,
where laboratory work is available. Langmuir--Hinshelwood as well as
Eley--Rideal surface chemistries are included in the model and their
relative contributions are highlighted. Analytic expressions are
derived for H$_2$, HD, and D$_2$ formation efficiencies for both type
of grains. Rate equations are tested against stochastic methods.} {As
expected, rate equations and stochastic methods diverge for grain
sizes lower than a critical value $a_{\rm crit}$. For grain sizes
below this critical value, \dm\ formation decreases to favour HD
formation. The formation efficiencies of \hm\ and \dm\ can be
calculated by adding a correction factor to the rate equations methods
(this factor is a simple exponential factor that becomes unity when
$a$ $>$ $a_{\rm crit}$).  We found that because of the presence of
chemisorbed sites, which can store atoms to form molecules up to high
grain temperatures, the formation efficiency of HD and D$_2$ is very
high compared to models where only physisorption sites are taken into
account.When considering a realistic distribution of dust grains, we
found that the formation rate of \hm \ and HD is enhanced by an order
of magnitude if small grains are taken into account. The formation of
\dm, on the other hand, is due to a contribution of small
($\le$100\AA) and big ($\ge$100\AA) grains, depending on the D/H
ratio, the grain temperature and the volume density. The processes
described in this paper, that allow a strong enhancement of the
deuterated forms of molecular hydrogen, could explain the high degree
of deuterium fractionation observed in protostellar environments.}{}{}


\titlerunning{\hm, HD and \dm}
\maketitle

\section{Introduction}

Formation of molecular hydrogen has been studied since decades but is
still the subject of polemics. Because of its importance to establish
the time scale of molecular cloud formation (e.g. Bergin et al. 2004),
as a coolant in low metallicity environments (Tegmark et al. 1997), or
as the primary ingredient for astrochemistry, the formation of \hm \,
occurring through grain surface reactions (Gould \& Salpeter 1963), is
the most studied surface reaction. However, there are considerable
uncertainties associated with this simple reaction mainly caused by
our ignorance concerning dust grain constitution (see the review by
Hollenbach \& Tielens 1999, where one also finds references to much of
the earlier work).  Because of this, recent investigations have
focused on the structure of dust grains, the interaction between grain
surfaces and accreting atoms, as well as the theoretical approach to
treat surface chemistry and to couple it with gas--phase chemistry
(Chang, Cuppen \& Herbst 2007). Interaction between H atoms and dust
grains has been extensively studied theoretically (Sha \& Jackson
2004, Sha \& Jackson 2002, Jeloaica \& Sidis 1999, Parneix \&
Brechignac 1998 ,Klose 1992) as well as experimentally (Pirronello et
al.1997a,b; 1999, Dulieu et al. 2005, Zecho et al. 2002, Hoenekaer et
al. 2003), allowing a better understanding of the morphology of dust
grains.

  The past discussion and also much recent work have centred on the
  fact that for the case (Langmuir Hinshelwood mechanism) where the
  two H atoms are physisorbed (attached to the surface by Van der
  Waals forces), one predicts efficient H$_{2}$ formation in a very
  small temperature range. This specificity is backed up by laboratory
  studies (Pironnello 1997 a, b; 1999).  Although the predicted
  temperature range for significant formation of \hm \ is close to the
  observed interstellar grain temperatures (Boulanger et al. 1996;
  Dwek et al. 1997; Rawlings et al. 2005) of around 15-18 ~K, it seems
  unlikely that this coincidence happens everywhere. Moreover, there
  are cases where the grain temperature is much higher and yet the
  H$_{2}$ formation rate appears high (Duley \& William 1984; Tielens
  \& Hollenbach 1985a, 1985b; Habart et al. 2004; Allers et al. 2005).
  Molecular hydrogen has been observed in the Universe under various
  physical conditions. In diffuse clouds, that typically have a
  density of 50 cm$^{-3}$, a gas temperature of $\sim\, 100$ K, and a
  dust temperature of 15 K, \hm\ forms with a rate of $1-3\times
  10^{-17}\, n_o\, n\left(H\right)$ cm$^{-3}$ s$^{-1}$ where $n_o$ is
  the total density and $n\left(H\right)$ is the density of H-atoms
  (Jura 1974; Hollenbach, Werner, \& Salpeter 1971). In molecular
  clouds and PDRs, the physical conditions at which \hm\ has been
  observed can cover a wide range of gas and grain temperatures
  (100K$\le$T$_{gas}$$\le$1000K; 10K$\le$T$_{dust}$$\le$100K). H$_2$
  formation rate varies from 3$\times 10^{-17}$ to 1.5$\times
  10^{-16}$ cm$^{-3}$ s$^{-1}$ for the PDRs associated with Orion Bar,
  NGC 2023, Chamaeleon, S140, IC 63 and Oph W (Habart et al. 2004,
  Allers et al. 2005). In Active galaxies, i.e. in galaxies hosting an
  active galactic nucleus (AGN), \hm\ emission originates from a
  region with a gas temperature $\le$2000K (in NGC 1068, Rigopoulou et
  al. 2002), and from a region with two dust components: one cold at
  15-30 K and one warm at 50-70 K (Barthel \& van Bemmel
  2003). Molecular hydrogen is also present in the early Universe, as
  testified by Damped Lyman Alpha systems (DLAs). Ledoux, Petitjean \&
  Srianand (2003) detected absorption lines of \hm\ toward several
  DLAs at high redshift (z$_{abs}$$\ge$1.8), showing that \hm\ could
  form even in low metallicity objects where the dust to gas ratio is
  as low as 1/20 to 1/4 of the Milky Way (Fall, Pei, \& McMahon
  1989). All these observations show that molecular hydrogen forms
  with a high rate under various physical conditions. This have led to
  various attempts to examine the circumstances under which the
  efficiency $\epsilon (T_{dust})$ could be close to 1 over an
  extended temperature range (Morisset et al. 2004, 2005; Cazaux \&
  Tielens 2002, 2004 ; Chang et al. 2005; Cuppen \& Herbst 2005).

The most recent approaches deal with the inhomogeneities of the
interstellar grain surfaces. The interactions between the atoms and
these surfaces are supposed to be only weak (Van der Waals) and are
distributed in energy. An approach adopted by Chang et al. (2005)
showed that on inhomogeneous surfaces with distributions of H--atom
diffusion barriers and (physisorption) binding energies, as well as on
mixed surfaces of olivine and carbon, the H$_2$ formation efficiency
remains large at dust temperatures relevant for diffuse clouds. More
recently, Cuppen \& Herbst (2005) considered in their Monte Carlo
simulations H$_2$ formation on {\it rough} surfaces (where, unlike
{\it flat} surfaces, there are height differences of at least several
monolayers), based on the structure of olivine and amorphous carbon
studied in laboratory experiments. Their results show an \hm\
formation efficiency large until $T_{\rm dust}$ $\sim$ 40~K depending
on the poorly known strengths of lateral bonds between H atoms and the
(rough) surface. In their work, the presence of surface irregularities
enhances the binding energy of hydrogen atoms but ignores a possible
barrier for the H atoms to enter in these sites, as it is the case for
strong interactions.

The approach of Cazaux and Tielens (2002,2004) considers that in
addition to physisorption sites on the grain, there are also sites
where H (or D) atoms can be chemisorbed with well depths of a large
fraction of an electron volt. The presence of these two interactions
between an atom and a surface has been calculated and measured by
several authors since decades (Physisorption: Ghio 1980; Pirronello et
al. 1997a, 1997b, 1999 and chemisorption: experimentally: G\"uttler et
al. 2004; Zecho et al 2002; Theoretically: Aronowitch \& Chang 1980;
Klose et al 1992; Fromherz et al. 1993; Jeloaica \& Sidis 1999; Sha \&
Jackson 2002; Sha \& Jackson 2005). By considering these two
interactions, one might have formation of H$_{2}$ due to association
of a physisorbed H atom with a chemisorbed H atom. Since the
chemisorbed H atom is more tightly bound will be less susceptible to
evaporation, one can thus maintain H$_{2}$ formation at grain
temperatures higher than the 15-18 K mentioned above.  Cazaux \&
Tielens (2004) examined this possibility and found that one could
indeed keep a high H$_{2}$ formation efficiency up to temperatures of
20-30 K with moderate efficiency up to several hundreds of
Kelvins. This seems more satisfactory though this efficiency depends
strongly (for T$_{dust} \ge$ 20-30K) on how easy an atom can chemisorb
on an interstellar grain, and therefore, on the nature of the barrier
against chemisorption. This uncertainties will be discussed in this
paper.

More laboratory work on olivine and carbonaceous grains will
definitely help in better constraining dust properties, but the actual
structure of interstellar dust grains cannot be completely defined if
model predictions cannot be tested against observations. As discussed
above, the H$_2$ formation process alone cannot be used for these
purposes, given that a large variety of models can reproduce the H$_2$
abundance observed in diffuse clouds. We need to go one step forward
and make other predictions besides the formation of H$_2$. The
simplest next step is to concentrate on the deuterated forms of
molecular hydrogen (HD and D$_2$) and this is the topic of the present
paper. This has traditionally been considered unimportant because HD
can form in the gas phase from H$_{2}$ via ion-molecule reactions (and
analogously D$_{2}$ from HD).  The gas phase route is expected to be
more efficient than direct formation on grains if the H$_{2}$
molecular fraction $f_{mol}\, = \, \frac{2n(H_{2})}{n_{H}}$ is larger
than roughly 0.1 (Watson 1973). In a future paper, we will examine
under which circumstances this is true and whether formation on dust
grains makes an appreciable contribution to the observed HD in diffuse
clouds. D$_{2}$ has not been observed to date and we study its
formation on grains as well as the possibility that it might be
observable. The present paper focuses on surface processes, whereas in
a futur paper, surface chemistry will be coupled with gas phase
chemistry and the model will be applied to different astrophysical
environments.

In section 2 of this paper, we describe the properties of dust surfaces, 
concentrating on Graphitic and amorphous carbon grains, for which laboratory 
work is available.  In Section 3 our model of surface chemistry is described,
including the H and D rates of evaporation and mobility, rate equations, 
formation efficiencies.  We also give some analytic approximations which
reasonably reproduce our computed efficiencies for
a limited range of grain temperature. In Section 4, we check the validity 
of the rate equation method against the Monte Carlo approach and compare with 
previous work.  A general discussion can be found in Section 5, whereas the 
main conclusions are listed in Section 6. 

\section{Properties of the dust surface}
\label{sproperties}

We concentrate here on Graphitic (Gr) and Amorphous Carbon (AC) grains
since Olivine surfaces have been experimentally studied only at low
temperatures (Pirronello et al. 1997a, 1997b), making a
characterisation of the strong interaction between the atoms and the
surface very doubtful (Cazaux \& Tielens 2002). Grains rich in carbon,
usually called carbonaceous in amorphous form, and graphitic in
crystalline form, have been the subject of a variety of experimental
and theoretical studies. Graphite surfaces have been favoured for
their simplest structures in order to perform ab-initio calculations
as well as experiments. Theoretically, studies agree in saying that
there is a barrier between physisorbed and chemisorbed sites of 0.2 eV
(Jeloaica \& Sidis 1999; Sha \& Jackson 2002), making the filling of
the chemisorbed sites very unlikely if the incoming atoms have low
energies. Experiments at high temperatures on graphite confirmed the
presence of a high barrier against chemisorption (Zecho et
al. 2002). Carbonaceous surfaces, on the other hand, present a very
small barrier against chemisorption. Experiments, performed by Menella
et al. (2001), show that for nano-sized carbon grains, low energies H
atoms can efficiently be bound to a the surface (C-H bonds 6$\%$),
showing that the barrier against chemisorption is small. Recent
experiments from Zecho (private communication) proved that amorphous carbon
surfaces present almost no barrier against
chemisorption. Theoretically, Sha \& Jackson (2004) determined a very
small activation barrier for the adsorption of H atoms on the edges of
Graphitic nanostructures, and proved that H atoms can easily be
chemisorbed on amorphous carbon surfaces.

\begin{figure*}
\includegraphics[width=1.0 \textwidth]{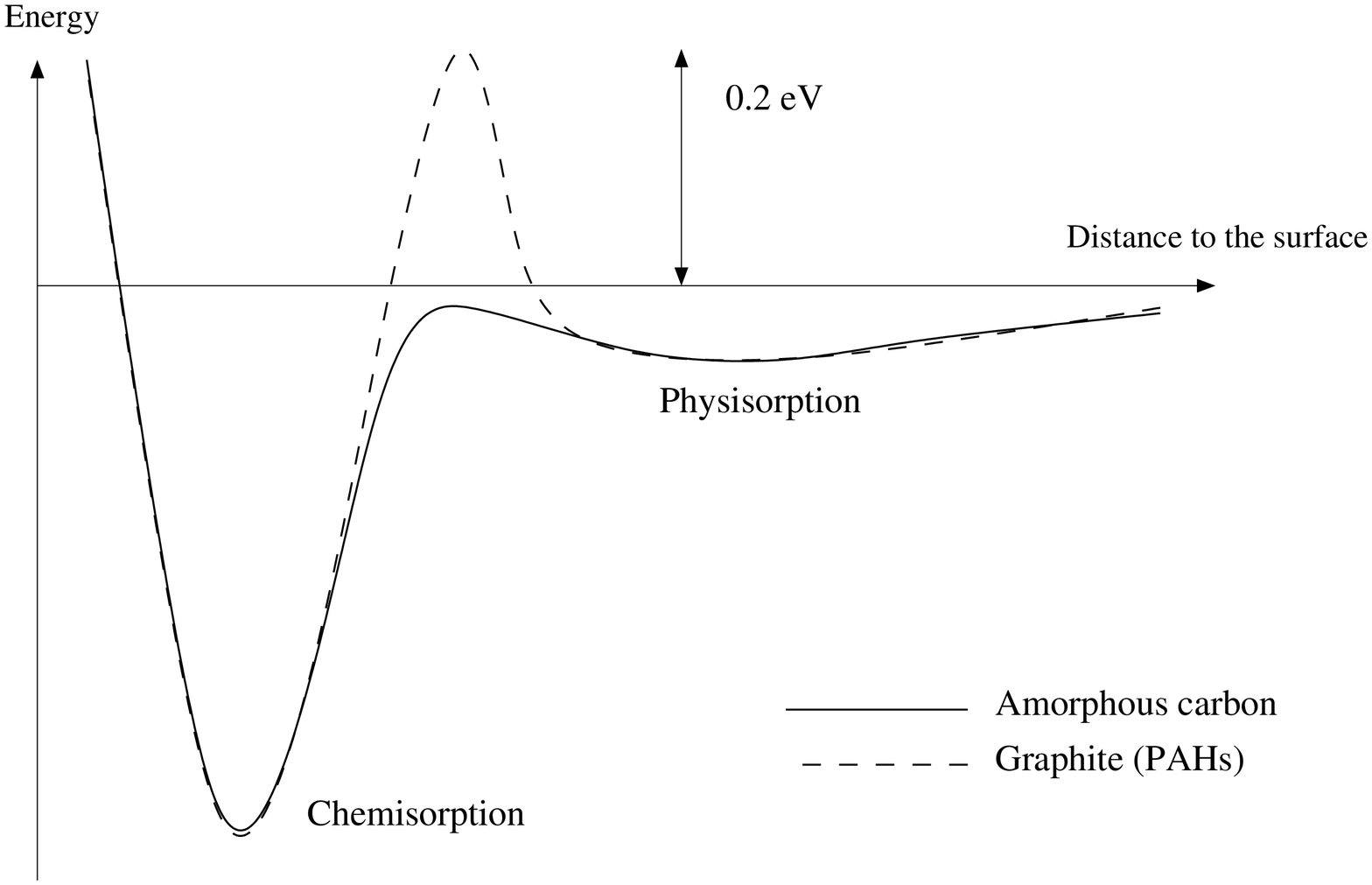}
\caption{Two possible types of carbonaceous grains.}
\label{barrier}
\end{figure*}

It seems that carbon grains in the diffuse medium are a mixture of
aliphatic and aromatic compounds (Pendleton \& Allamandola 2002). Big
dust grains are thought to be under the form of carbonaceous grains,
whereas small dust grains are under the form of PAHs. In this work, we
consider two types of carbon grains as illustrated in
Fig~\ref{barrier}: (1) Small grains ($\le$ 100\AA), also called PAHs,
have a surface similar to graphite. This surface present a barrier
against chemisorption of 0.2 eV height and suppress the formation of
\hm\ at intermediate and high temperatures. (2) Bigger grains, have a
surface similar to Amorphous Carbon surface. This surface presents a
small (or none) barrier against chemisorption.

\section{Surface chemistry}
\label{ssurface}

The formation of \hm\ and its deuterated forms on cold grains (T$_{dust}
\le$20K) has been discussed by many authors, and is still a
controversial subject. While the Langmuir--Hinshelwood kinetics -- the
association of two atoms moving on the surface -- is one of the main
processes for the formation of these molecules at low grain
temperatures (Pirronello et al. 1997a, 1997b, 1999; Katz et al. 1999;
Chang et al. 2005; Cuppen \& Herbst 2005), the Eley--Rideal mechanism
-- the association of an atom from the gas phase with a physisorbed
atom -- seems not negligible (Morrisset et al. 2003). We consider in
our model these two mechanisms, and discuss in Sect.~\ref{lh_er} their
relative importance.

\subsection{Interaction between H, D atoms and the dust surface}
An atom from the gas phase hits the grain surface with a certain
energy $kT_{gas}$ and, if stuck on the grain, it can either become
physisorbed or chemisorbed. The probability for an atom to directly
arrive in a chemisorbed site is given by the transmission coefficient
(to pass the barrier against chemisorption) integrated over an energy
range described by the Boltzmann law (for details see Cazaux \&
Tielens 2004). We call this probability T$_{pc}$(H$_{gas}$) and
T$_{pc}$(D$_{gas}$) for H and D atoms, respectively. Once on the
surface, the atom can go from its site to another site with a rate
$\alpha_{ij}$, where $i$ is the initial site, and $j$ is the next site
which can be occupied. These rates, which describe the mobility of
atoms on the surface, are calculated similarly to the
T$_{pc}$(H$_{gas}$) and T$_{pc}$(D$_{gas}$), but they are multiplied
by the oscillation factor of the atoms. Also, the Boltzmann energy
distribution of an atom on a grain is centered on T$_{dust}$, while it
is centered on T$_{gas}$ for an atom in the gas phase. The atom on the
grain can evaporate with a rate $\beta(H_P)$ and $\beta(D_P)$ for
physisorbed H and D atoms, and $\beta(H_C)$ and $\beta(D_C)$ for
chemisorbed H and D atoms. Figure~\ref{surf} shows the various rates
adopted in our model as a function of dust temperature and for the two
different types of grains considered (see
Sect.~\ref{sproperties}). $\alpha_{\rm pc}$ and $\alpha_{\rm pp}$ are
the mobility rates of H or D to move from a physisorbed to a
chemisorbed site and between two physisorbed sites, respectively.
$\beta$ is the evaporation rate for physisorbed deuterium and
hydrogen, exponentially dependent on the temperature. T$_{pc}$ is the
fraction of the H and D atoms coming from the gas phase, with a
temperature of 100K, that directly chemisorb (1 $\%$ on an amorphous
carbon surface, and 0.02 $\%$ on graphitic surfaces). The first thing
to note in the figure is the huge difference between $\alpha_{\rm pc}$
for Gr and AC grains at $T_{\rm dust}$ $>$ 20~K, which shows weak
dependence with dust temperature when tunnelling dominates ($T_{\rm
dust}$ $<$ 20~K for AC and $<$ 100~K for Gr), whereas the dependence
becomes exponential when thermal hopping dominates. We also note the
steep rise of $\alpha_{\rm pp}$(D) at $T_{\rm dust}$ $>$ 11~K and
$\alpha_{\rm pp}$(H) at $T_{\rm dust}$ $>$ 14~K for Gr and AC grains.

\subsection{Rate equations}
The model described in Cazaux \& Tielens (2004) is reconsidered here,
with the addition of deuterated species. \hm, HD and \dm\ formation
mechanisms on grain surfaces are studied with the rate equation method
based on the following assumptions: \\ 1) The interaction atom/surface
can be either weak (Van der Waals interaction, also called
physisorption) or strong (covalent bound also called chemisorption),
with the assumption that for each physisorbed site, there is a
chemisorbed site.\\ 2) An atom on the grain moves from one site to
another by tunnelling effect or thermal hopping, according to its
energy.\\3) Two atoms can associate to form a molecule following the
Langmuir--Hinshelwood and Eley--Rideal mechanisms.

In this study, we add deuterium in our rate equation model and follow
the population of physisorbed H (H$_P$) and D (D$_P$), chemisorbed H
(H$_C$) and D (D$_C$), and the \hm, HD and \dm\ molecules, which are
physisorbed. The rate equations are written as follow:
 
\begin{eqnarray}
\dot{H}_{P}&=&F_H(1-T_{pc}H_{gas})(1-H_{P}-D_{P}-{HD}-{H_{2}}-{D_{2}})\nonumber\\
&&-F_H(1-T_{pc}H_{gas})H_{P}-F_D(1-T_{pc}D_{gas})H_{P}\nonumber\\
&&-\alpha_{pc}(H)H_{P}-2\alpha_{pp}(H){H_{P}}^2-\alpha_{pp}(H)H_{P}D_{P}-\alpha_{pp}(D)D_{P}H_{P}\nonumber\\
&&-\beta_{H_P}H_{P}
\end{eqnarray}

\begin{eqnarray}
\dot{D}_{P}&=&F_D(1-T_{pc}D_{gas})(1-H_{P}-D_{P}-{HD}-{H_{2}}-{D_{2}})\nonumber\\ &&-F_D(1-T_{pc}D_{gas})D_{P}-F_H(1-T_{pc}H_{gas})D_{P}\nonumber\\
&&-\alpha_{pc}(D)D_{P}-2\alpha_{pp}(D){D_{P}}^2-\alpha_{pp}(D)D_{P}H_{P}-\alpha_{pp}(H)H_{P}D_{P}\nonumber\\
&&-\beta_{D_P}D_{P}
\end{eqnarray}
These two equations describe the fractional population of H and D
atoms in physisorbed sites, when a flux F$_H$ and F$_D$ of H and D
atoms is sent on the surface. The first two lines of these equations
describe the fraction of H and D atoms coming from the gas phase that
physisorb on the grain. A fraction T$_{pc}H_{gas}$ and
T$_{pc}D_{gas}$, of H and D atoms goes directly in chemisorbed
sites. These fractions are the amount of H and D atoms, coming from
the gas phase with an energy T$_{gas}$, that can cross the barrier
against chemisorption.  Also, a part of the incoming atoms going to
physisorb arrive in an already filled physisorbed site and can either
form directly a molecule (Eley-Rideal mechanism, 2$^{nd}$ lines of
eq. 1 and 2) if the site is occupied by an atom, or bounce back to the
gas phase if the site is occupied by a molecule. The third line of
these equations are dealing with Langmuir kinetic processes where the
atoms diffuse on the surface from a site i to a site j with a rate
$\alpha_{ij}(H)$ and $\alpha_{ij}(D)$. On these third lines, the first
terms are the rate of atoms going to chemisorbed sites, the second,
third and fourth terms are the association of the physisorbed atoms
with other physisorbed atoms and the last term is the evaporation rate
of the physisorbed atoms.\\  The equations describing the
fractional population of H and D atoms in chemisorbed sites are:
\begin{eqnarray}
\dot{H}_{C}&=&F_H\ T_{pc}H_{gas}\ (1-H_{C}-D_{C})-F_H\ T_{pc}H_{gas}\ H_{C}-F_D\ T_{pc}D_{gas}\ H_{C}\nonumber\\
&&+\alpha_{pc}(H)H_{P}(1-H_{C}-D_{C})-\alpha_{pc}(H)H_{P}H_{C}-\alpha_{pc}(D)D_{P}H_{C}\nonumber\\
&&-\beta_{H_C}H_{C}
\end{eqnarray}
\begin{eqnarray}
\dot{D}_{C}&=&F_D\ T_{pc}D_{gas}\ (1-D_{C}-H_{C})-F_D\ T_{pc}D_{gas}\ D_{C}-F_H\ T_{pc}H_{gas}\ D_{C}\nonumber\\
&&+\alpha_{pc}(D)D_{P}(1-D_{C}-H_{C})-\alpha_{pc}(D)D_{P}D_{C}-\alpha_{pc}(H)H_{P}D_{C}\nonumber\\
&&-\beta_{D_C}D_{C}
\end{eqnarray}
The first lines represent the fraction of the H and D atoms that cross
the barrier against chemisorption and populate directly the
chemisorbed sites (with probability $T_{pc}$(H) and $T_{pc}$(D),
respectively). One part of these atoms arrive in some already occupied
sites and form a molecule (Eley-Rideal mechanism, 2$^{nd}$ and
3$^{rd}$ terms of the first lines). In the second lines, the first
terms account for physisorbed atoms arriving in empty chemisorbed
sites, the second and third terms for the association of incoming
physisorbed atoms with chemisorbed atoms, and the last term accounts
for the evaporation of chemisorbed atoms.

The fractional population of \hm, D$_2$ and HD on grain surfaces 
is described by the following three equations:
\begin{eqnarray}
\dot{H}_{2}&=&+\mu(\alpha_{pp}(H){H_{P}}^{2}+\alpha_{pc}(H)H_{P}H_{C})-\beta_{H_{2}}H_{2}
\end{eqnarray}
\begin{eqnarray}
\dot{D}_{2}&=&+\mu(\alpha_{pp}(D){D_{P}}^{2}+\alpha_{pc}(D)D_{P}D_{C})-\beta_{D_{2}}D_{2}
\end{eqnarray}
\begin{eqnarray}
\dot{HD}&=&+\mu(\alpha_{pp}(H)H_{P}D_{P}+\alpha_{pp}(D)D_{P}H_{P}+\alpha_{pc}(H)H_{P}D_{C}\nonumber\\&&+\alpha_{pc}(D)D_{P}H_{C})-\beta_{HD}HD
\end{eqnarray}
When a molecule is formed, because of the energy released during
formation, a fraction $\mu$ stays on the grain surface, while a
fraction 1-$\mu$ spontaneously desorbs in the gas phase. The value of
$\mu$ has been estimated from experiments (Katz et al. 1999; Cazaux \&
Tielens 2002), and is considered to be identical for the different
species (based on the experiments of HD desorption from Pirronello et
al. 1997). The terms multiplied by $\mu$ describe the total formation
rate of species remaining on the surface, whereas the last terms
correspond to their evaporation rate.{We consider in our
approach that the molecules formed through the Eley-Rideal mechanism
directly desorb in the gas phase. In the next section, we show that
this approximation is reasonnable for grain temperatures higher than
10K.}

In order to estimate the coverage of the different atoms and molecules
on the grain, one needs to solve the 7 rate equations above, which are
coupled. We point out here that the rate equation approach is valid
when grains are covered by 1 or more of each species considered. This
point is discussed in Sect.~\ref{svalidity} of this paper.

\subsection{Formation rates and formation efficiencies}
\label{smoldust}
The formation rate on dust surface of \hm, HD and \dm\ that are
released in the gas phase can be written as:

\begin{eqnarray}
R_d({H}_{2})&=&F_H(1-T_{pc}H_{gas})H_{P}+F_H\ T_{pc}H_{gas}\ H_{C}\nonumber\\
&&+(1-\mu)(\alpha_{pp}(H){H_{P}}^{2}+\alpha_{pc}(H)H_{P}H_{C})+\beta_{H_{2}}H_{2}
\end{eqnarray}
\begin{eqnarray}
R_d({D}_{2})&=&F_D (1-T_{pc}D_{gas}) D_{P}+F_D\ T_{pc}D_{gas}\ D_{C}\nonumber\\
&&+(1-\mu)(\alpha_{pp}(D){D_{P}}^{2}+\alpha_{pc}(D)D_{P}D_{C})+\beta_{D_{2}}D_{2} 
\end{eqnarray}
\begin{eqnarray}
R_d({HD})&=&F_H(1-T_{pc}H_{gas}) D_{P} +F_H\ T_{pc}H_{gas}\ D_{C}\nonumber\\
&&+F_D\ (1-T_{pc}D_{gas})\ H_{P} + F_D\ T_{pc}D_{gas}\ H_{C}\nonumber\\
&&+(1-\mu)(\alpha_{pp}(H)H_{P}D_{P}+\alpha_{pp}(D)D_{P}H_{P}\nonumber\\
&&+\alpha_{pc}(H)H_{P}D_{C}+\alpha_{pc}(D)D_{P}H_{C})+\beta_{HD}HD 
\end{eqnarray} 
In these expressions, the terms with F$_H$ and F$_D$ represent the
association of gas phase atoms with adsorbed atoms to form molecules
via the Eley--Rideal mechanism. The other terms involve the
association of two adsorbed atoms through the Langmuir-Hinshelwood
kinetic. The formation efficiencies of the different molecules are:
$\epsilon({H_2})=\frac{2\times R_d({H_2})}{F_H}$,
$\epsilon({D_2})=\frac{2\times R_d({D_2})}{F_D}$ and
$\epsilon({HD})=\frac{R_d({HD})}{F_D}$.

\subsubsection{Eley--Rideal vs Langmuir--Hinshelwood}
\label{lh_er}

The efficiency of formation of \hm, HD and \dm\ for graphitic and
amorphous carbon surfaces are shown in Fig.~\ref{Eff}, for a density
of H atoms of 100 atoms cm$^{-3}$, a gas temperature of 100K, and a
D/H ratio of 2 10$^{-5}$. The different mechanisms involved in the
formation of these molecules are presented. The Langmuir--Hinshelwood
mechanism (LH) is the dominant process at any dust temperatures for
amorphous carbon grains and at low grain temperatures (T$_{dust} \le$
25K) for graphitic grains. Then, only in the case of graphitic grains,
and at higher grain temperatures (T$_{dust} \ge$ 25K), the Eley-Rideal
(ER) mechanism becomes important because most of the physisorbed atoms
evaporate before populating the chemisorbed sites. The formation of
molecules is then assured by the association of those atoms that
chemisorb with the atoms from the gas phase that cross the barrier
against chemisorption.

{We consider in our model that molecules formed through the
Eley--Rideal mechanisms are directly released into the gas
phase. There is actually no experimental prove of such a behaviour,
but because the Eley-Rideal mechanism is efficient for grain
temperatures higher than 25K, if some newly formed molecules remain on
the surface, they will evaporate immediately. Therefore, considering
$\mu =0$ for the Eley-Rideal mechanism is actually a reasonable
approximation. In the case of very cold grains $\le$ 9K, this
approximation can not be applied and a factor $\mu$ $\neq 0$ should be
considered.}

\begin{figure*} 
\includegraphics{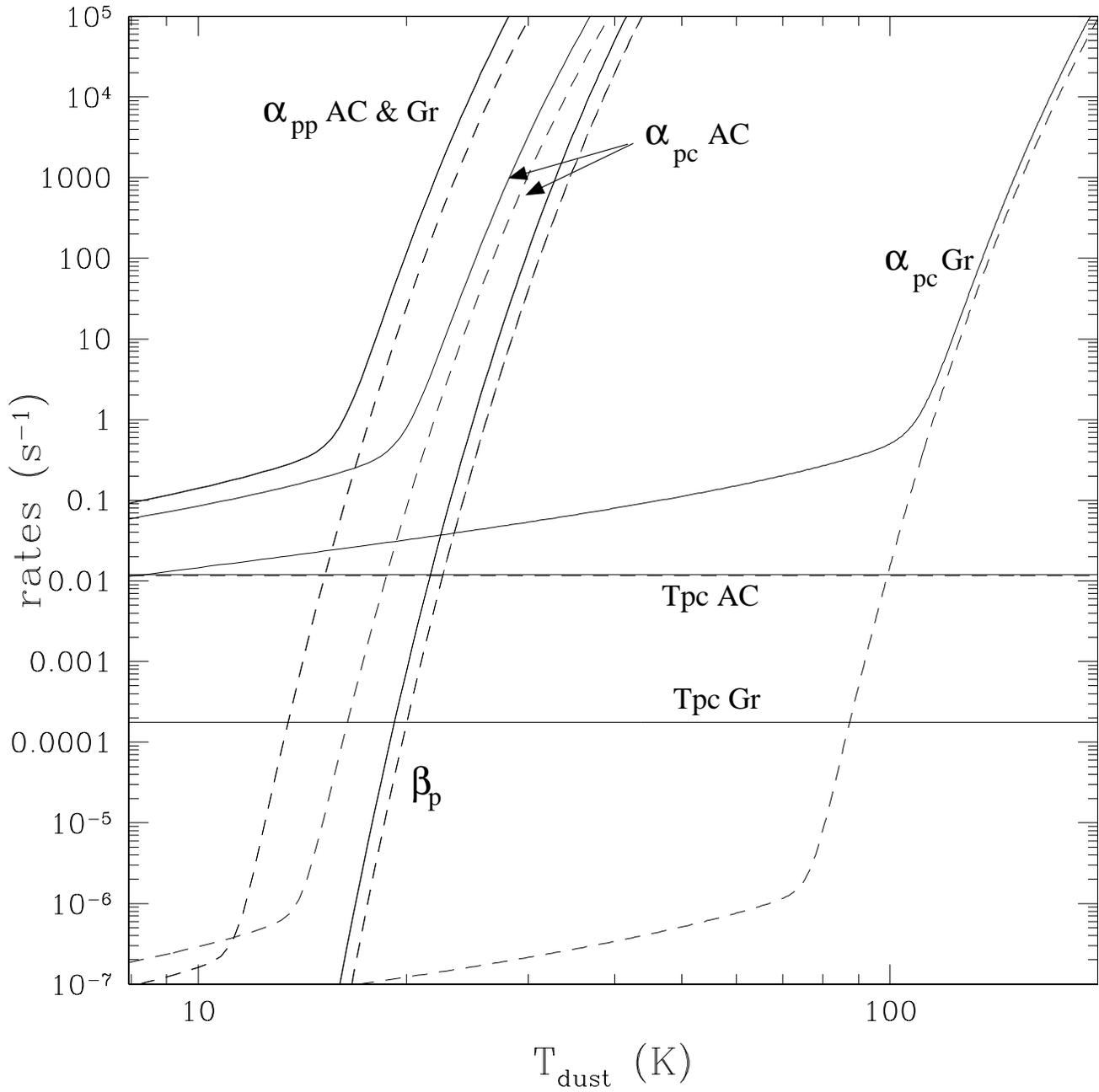}
\caption{Mobility and evaporation rate of the H (solid lines) and D
(dashed lines) atoms on the graphitic surfaces (Gr) and amorphous
carbon surfaces (AC).}
\label{surf}
\end{figure*}

A striking result is the enhanced \dm\ formation for a very narrow
range of grain temperatures, while \hm \ and HD form efficiently for a
much broader range. This \dm\ enhancement is much more important on
graphitic surfaces than on amorphous carbon surfaces. Indeed, in the
case of graphitic surfaces, at low surface temperatures, the high and
narrow barrier against chemisorption allows H atoms to chemisorb
through tunnelling, whereas most of the D atoms stay physisorbed. The
H atoms, imprisoned in chemisorbed sites, free the physisorbed sites
and let the physisorbed D atoms meet other physisorbed D atoms. This
process makes the formation of \dm\ very efficient (for T$_{dust}$
$\le$ 20K), but for higher grain temperatures, the physisorbed D
evaporate and no chemisorbed D can insure the formation of
\dm. Therefore the \dm\ efficiency drops for grain temperatures higher
than 20K. In the case of carbonaceous grains, the barrier against
chemisorption is smaller, and H and D atoms present similar mobilities
(see Fig.~\ref{surf}) making the physisorbed sites also populated by H
atoms. The formation of \dm\ is slightly enhanced around T$_{dust}$
$\sim$ 15K and follows the same behaviour than \hm\ and HD at higher
grain temperatures.

The grain temperature at which \dm\ formation reaches its maximum also
varies with the type of grain.  In fact, the formation of \dm\ depends
on the amount of physisorbed D atoms, which decreases with the
physisorption--chemisorption mobility and evaporation rates,
$\alpha_{pc}(D)$ and $\beta_{D_P}$, both strongly dependent on $T_{\rm
dust}$ (see Fig.~\ref{surf}). For graphitic grains, at the highest
\dm\ formation efficiency, the physisorbed D atoms evaporate more
easily than they move into a chemisorbed site ($\alpha_{pc}(D) \le
\beta_{D_P}$). For amorphous carbon grains, on the other hand, the
physisorbed D atoms disappear by getting trapped in chemisorbed sites
(for low H densities $\sim$ 1 cm$^{-3}$) or by associating with a
physisorbed H atom (only for high H densities $\sim$ 10$^4$
cm$^{-3}$). Therefore, the decrease of \dm\ efficiency, due to the
disappearance of physisorbed D, occurs at lower grain temperatures for
Gr than for AC grains.

\begin{figure*}
\includegraphics[width=0.5\textwidth]{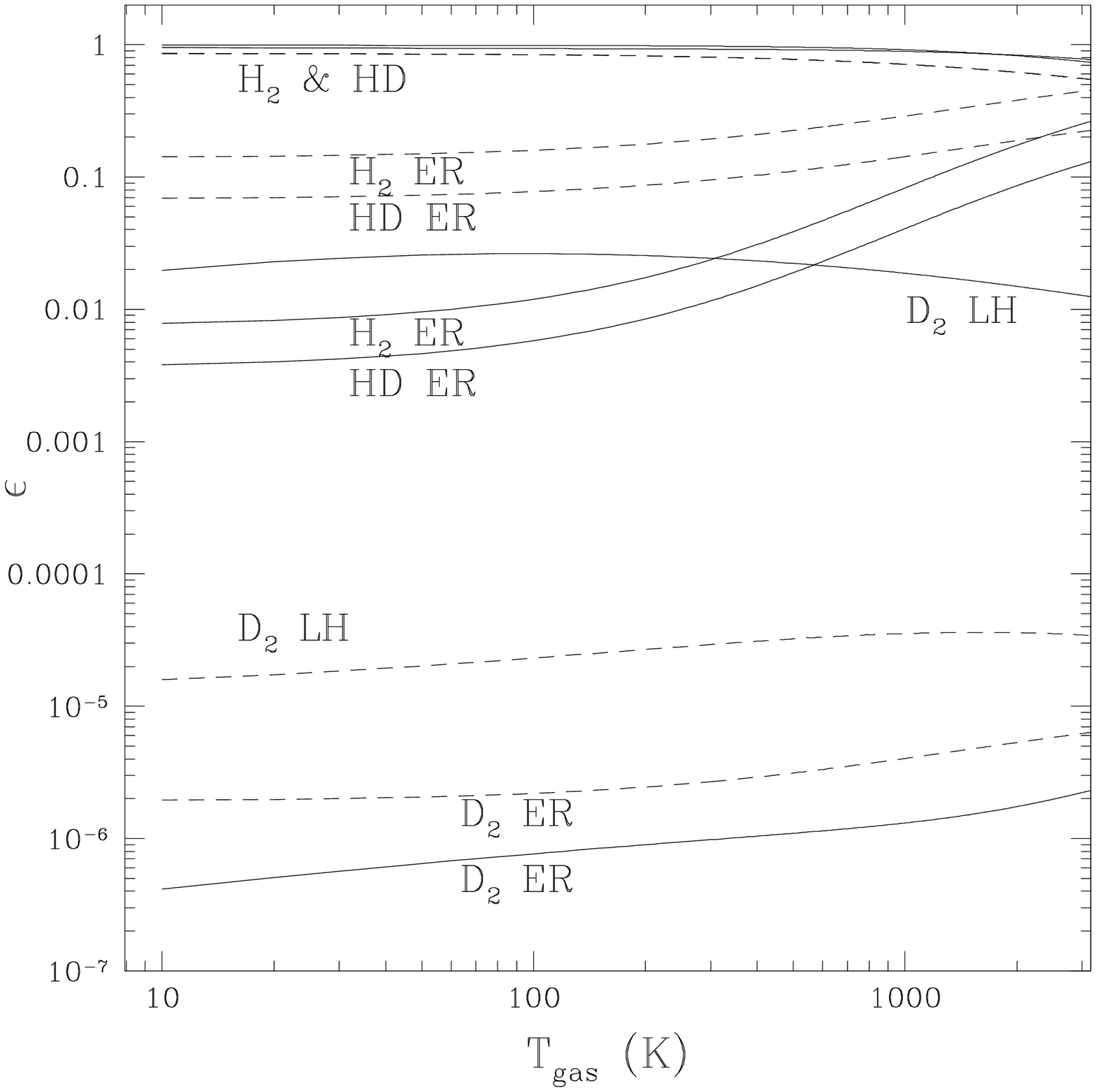}
\hspace{0.5cm}
\includegraphics[width=0.5\textwidth]{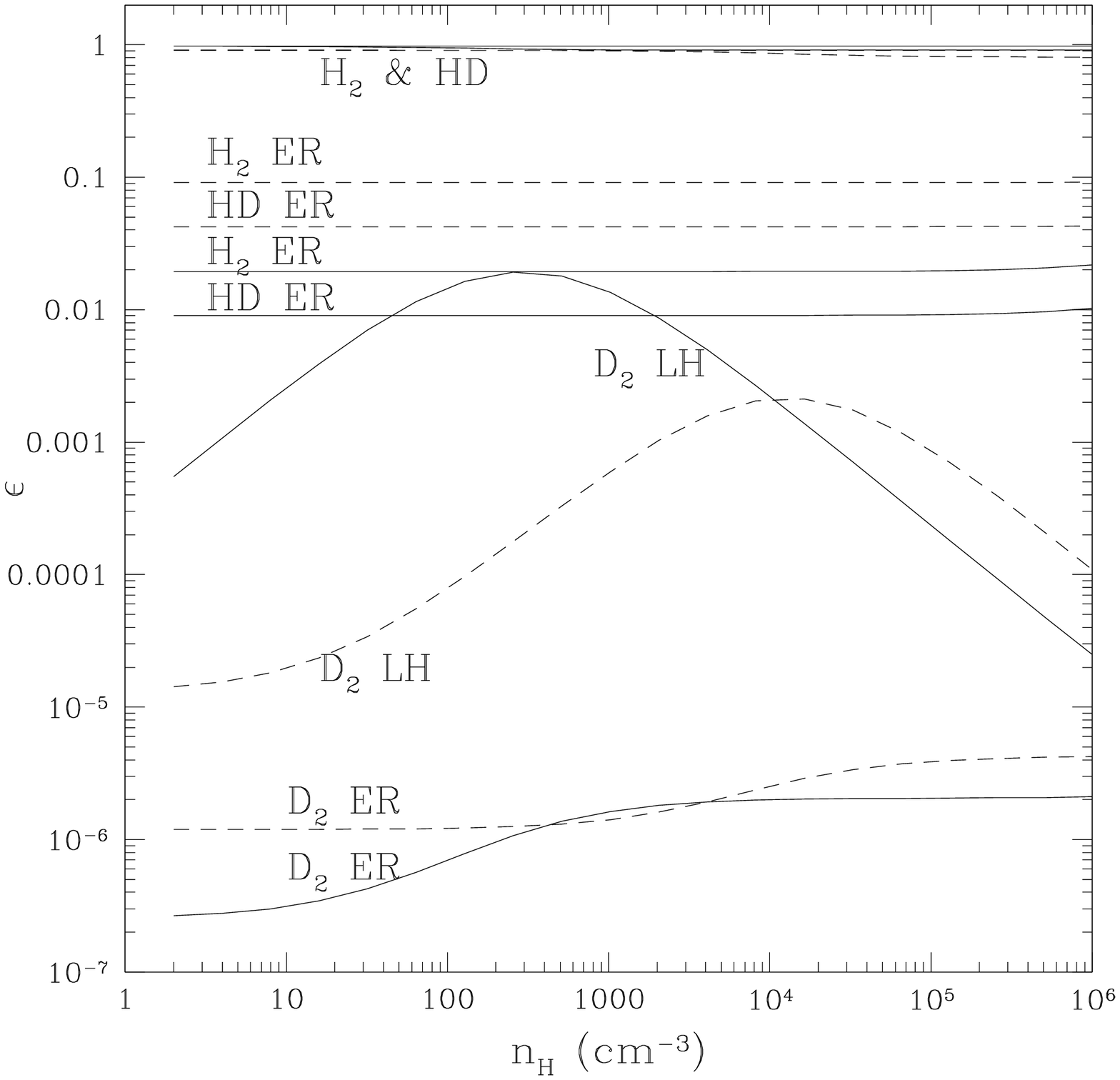}
\caption{\hm , HD and \dm\ formation efficiencies on graphitic (solid
lines) and amorphous carbon surfaces (dotted dashed lines).  Left:
Efficiencies as a function of the gas temperature, with a density of
100 cm$^{-3}$. Right: Efficiencies as a function of the density nH,
with a gas temperature set at 100 K.}
\label{thnh}
\end{figure*}

The effect of the variation of gas temperature and H density is shown
in Fig.~\ref{thnh}, with a grain temperature set at 15K. The gas
temperature does not have a big impact on the formation of \hm, HD
and \dm\ with the Langmuir--Hinshelwood mechanism. On the other hand,
the gas temperature strongly increases the formation of the different
molecules through the Eley-Rideal mechanism because the atoms can
cross more easily the barrier against chemisorption.

As shown in Fig.~\ref{thnh}, the \hm\ and HD formation efficiencies
are constant with H density, while \dm\ formation efficiency is
sensitive to its variation. The Langmuir--Hinshelwood kinetic is
strongly affected by the density of H atoms. Indeed, for low H
densities, the physisorbed sites are mainly populated with D atoms,
and \dm\ formation is enhanced, while for high H densities,
physisorbed sites are also populated by H atoms, making easier the
formation of an HD molecule then of a \dm\ molecule. This effect is
very different for the two types of grain since the barrier between
physisorption and chemisorption sets the migration time of the atoms
to go to a chemisorbed site, 1/$\alpha_{pc}(H)$ and
1/$\alpha_{pc}(D)$. If the physisorbed D atoms go to a chemisorbed
site easier than they evaporate ($\alpha_{pc}(D) \ge \beta_{D_P} $),
or than they meet a physisorbed H atom ($\alpha_{pc}(D) \ge
\frac{\alpha_{pp}(H)}{\alpha_{pc}(H)}F_H$ where this expression
represents the flux of H atoms in physisorbed sites), then D atoms
transit to chemisorbed sites, and less \dm\ forms. If, on the other
hand, the physisorbed D atoms have more chance to meet another
physisorbed D than to chemisorb
($\frac{\alpha_{pp}(D)}{\alpha_{pc}(D)}F_D$ becomes $ \ge
\alpha_{pc}(D)$) or to meet a physisorbed H
($\frac{\alpha_{pp}(D)}{\alpha_{pc}(D)}F_D$ becomes $ \ge
\frac{\alpha_{pp}(H)}{\alpha_{pc}(H)}F_H$), then \dm\ is formed more
efficiently. For the two types of grains considered here, the \dm\
efficiency reaches its maximum at different H densities (nH=100 atoms
cm$^{-3}$ for Gr grains and nH=10$^4$ atoms cm$^{-3}$ for AC grains).

\begin{figure*}
\includegraphics[width=0.5\textwidth]{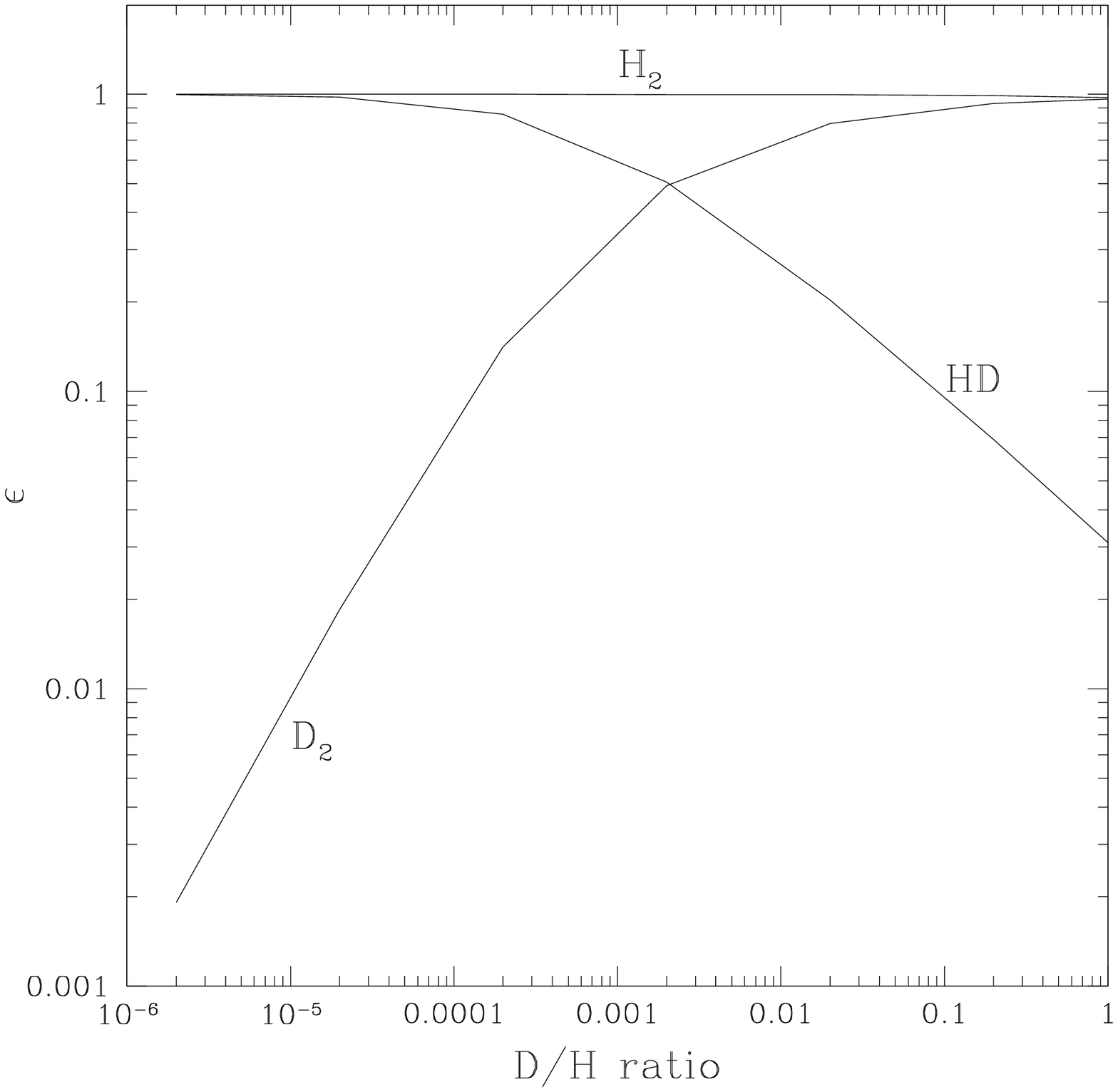}
\hspace{0.5cm}
\includegraphics[width=0.5\textwidth]{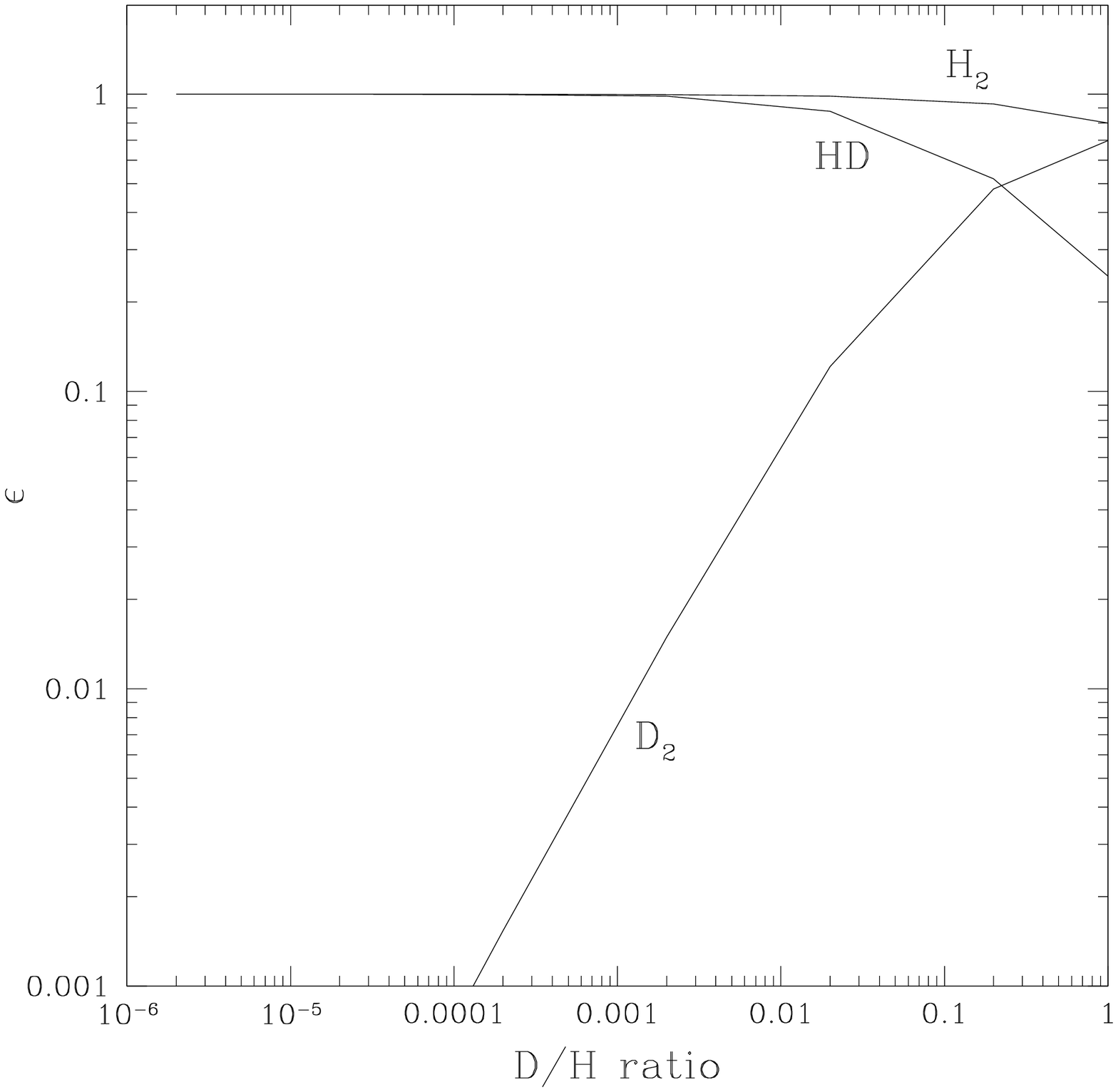}
\caption{HD and \dm\ formation efficiencies on graphitic (left panel)
and amorphous carbon surfaces (right panel), as a function of the D/H
ratio. The grain temperature is set at 15 K, the gas temperature at 100
K and the density at 100 cm$^{-3}$.}
\label{dh}
\end{figure*}

The effect of the variation of the D/H ratio on the formation
efficiencies of the different molecules is reported in Fig.~\ref{dh},
with a grain temperature set at 15K, a gas temperature at 100K and a
density of 100 atoms cm$^{-3}$. \hm\ does not depend on the D/H ratio,
while HD and \dm\ efficiencies show very different behaviours for Gr
and AC grains. As discussed before, Gr grains, which have a very high
barrier against chemisorption, segregate the H and D atoms by having
most of its H chemisorbed and D physisorbed. This segregation favours
deuteration.

\subsubsection{Analytic expressions}

Between dust temperatures of 10K and 25 K, as discussed previously,
the Eley-Rideal mechanism can be neglected. The formation efficiency
of \hm, HD and \dm\ are calculated by setting the rate equations to
zero (steady state conditions), and can be approximated as:
\begin{equation}
\epsilon_{H_2}= 2 \times \frac{(\alpha_{pp}(H) H_P^2 + \alpha_{pc}(H) H_P H_C)}{F_H}
\end{equation}

\begin{equation}
\epsilon_{D_2}= 2 \times \frac{ (\alpha_{pp}(D) D_P^2 + \alpha_{pc}(H) H_P D_C)}{F_D}
\end{equation}

\begin{equation}
\epsilon_{HD}=\frac{\alpha_{pp}(H)+\alpha_{pp}(D)) H_P D_P+ \alpha_{pc}(H) H_P D_C+ \alpha_{pc}(D) D_P H_C}{F_D}
\end{equation}

These equations show that the formation of \hm\ occurs mainly through
the association of a physisorbed and a chemisorbed atom. HD can be
formed through the association of 2 physisorbed atoms, when the H
atoms populate the physisorbed sites (high H densities of the medium,
and low grain temperatures) or through the association of physisorbed
and chemisorbed atoms. In the same way, \dm\ can be form through the
association of two physisorbed atoms when physisorbed sites are mostly
populated by D atoms, or through the association of a physisorbed and
a chemisorbed D atom. The population of physisorbed H (H$_P$) and D
(D$_P$), as well as chemisorbed H (H$_C$) and D (D$_C$) and \hm, HD
and \dm\ are determined in steady state conditions as:

\begin{equation}
H_P=\frac{F_H}{\alpha_{pc}(H)+\beta_{H_P}}
\end{equation}
\begin{equation}
D_P=\frac{\alpha_{pc}(D)+\alpha_{pp}(H) H_P+ \sqrt{(\alpha_{pc}(D)+\alpha_{pp}(H) H_P)^2 + 4 F_D \alpha_{pp}(D)}}{2\alpha_{pp}(D)}
\end{equation}
\begin{equation}
H_C=\frac{1}{2}
\end{equation}
\begin{equation}
D_C=\frac{\alpha_{pc}(D) D_P}{2\alpha_{pc}(H)
H_P}
\end{equation}
\begin{equation}
H_2=\frac{\mu \alpha_{pc}(H)H_P H_C}{\beta_{H_2}}
\end{equation}
\begin{equation}
HD=\frac{\mu (\alpha_{pp}(H)+\alpha_{pp}(D)) H_P D_P+ \alpha_{pc}(H)
H_P D_C+ \alpha_{pc}(D) D_P H_C)}{\beta_{HD}}
\end{equation}
\begin{equation}
D_2=\frac{\mu(\alpha_{pp}(D) D_P^2 + \alpha_{pc}(D) D_P
D_C)}{\beta_{D_2}}  
\end{equation}

{Half of the chemisorbed sites are occupied by hydrogen atoms
because at low temperatures, the rate of physisorbed atoms to become
chemisorbed is higher than their evaporation rate (as shown in
fig.~\ref{surf}). Therefore physisorbed atoms, after visiting a number
of $\frac{\alpha_{pp}}{\alpha_{pc}}$ of physisorbed sites, will become
trapped in chemisorbed sites. Above a certain temperature (20K for
graphite and 50 K for amorphous carbon), this sites are filled through
direct chemisorption.}

The formation efficiencies of the different species, for grain
temperatures between 10 and 25 K, can be written:

\begin{equation}
\epsilon_{H_2}= \left[\frac{\alpha_{pc}(H) + \beta_{H_P}}{\alpha_{pc}(H)}\right]
\end{equation}

\begin{equation}
\epsilon_{D_2}= 2 \frac{D_P^2}{F(D)} \left[\alpha_{pp}(D)+\frac{\alpha_{pc}(D)^2}{2 F(H)}\right]
\end{equation}

\begin{equation}
\epsilon_{HD}=D_P \left[\frac{(\alpha_{pp}(H)+\alpha_{pp}(D))H_P+\alpha_{pc}(D)}{F_D}\right]
\end{equation}

\begin{figure*}
\includegraphics[width=0.5\textwidth]{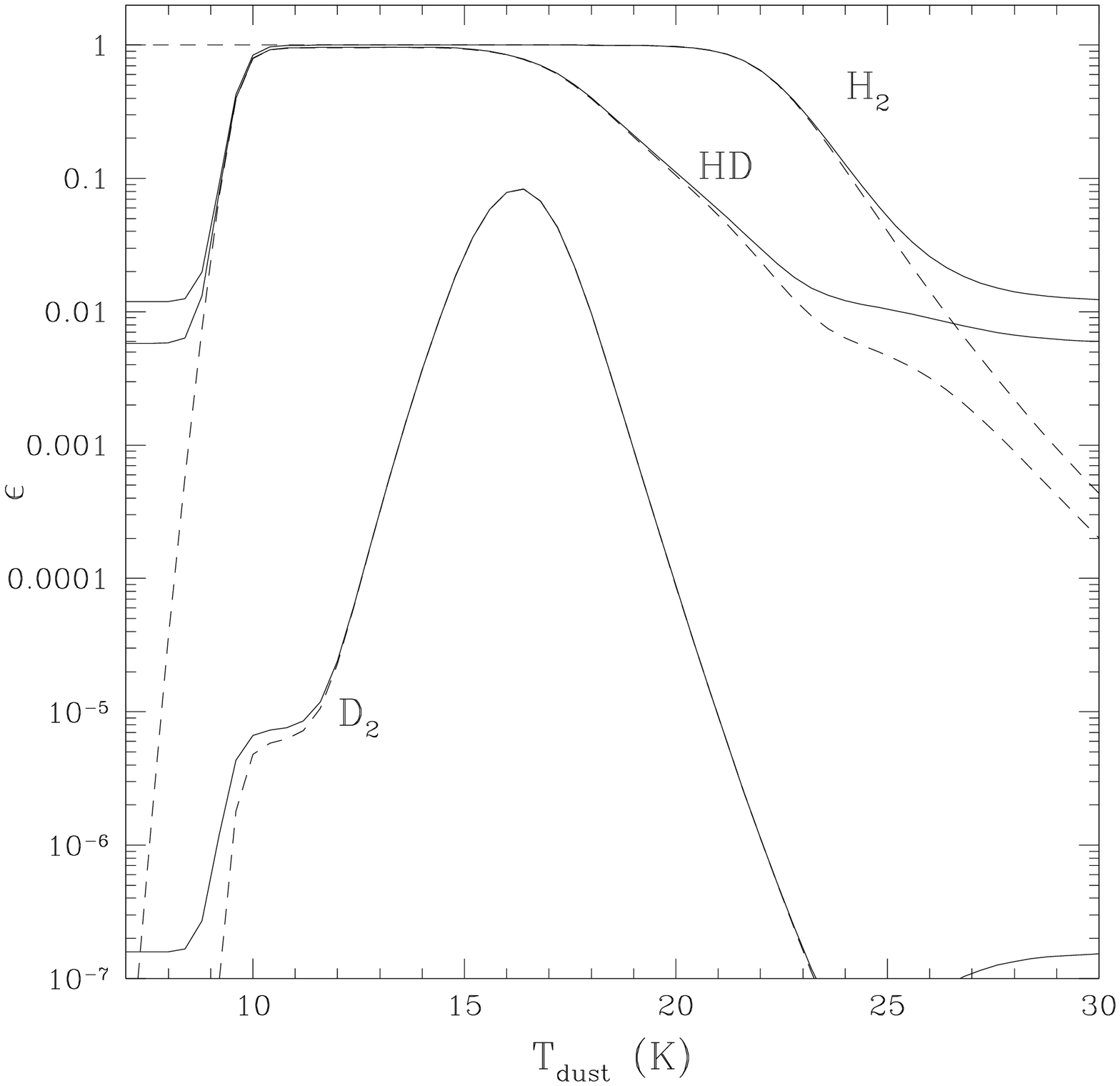}
\hspace{0.5cm}
\includegraphics[width=0.5\textwidth]{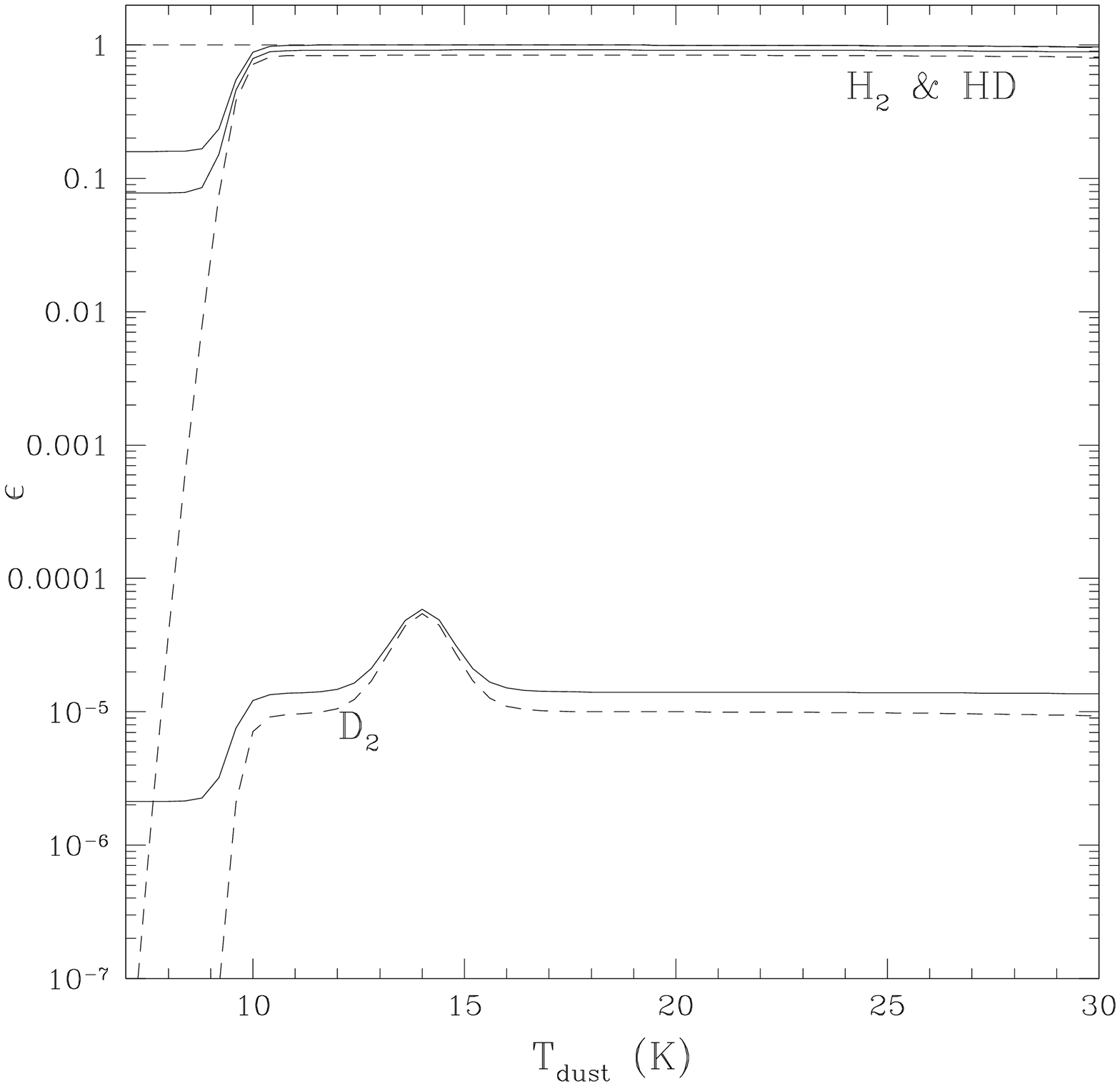}
\caption{\hm, HD and \dm\ formation efficiencies on graphitic (left)
and amorphous carbon (right) surfaces compared to our approximations
(dashed lines).}
\label{apx}
\end{figure*}

In Fig.~\ref{apx}, we show the validity of these approximations with
our model. As discussed above, we consider a range of grains
temperatures (10K $\le$ T$_{dust}$ $\le$ 25K) in which the Eley
Rideal mechanism can be neglected. Because we are using a rate
equation approach to calculate the formation rate of the different
molecules, we need to define for which circumstances our approach is
not valid. Indeed, rate equations cannot be used if the grain is
covered by less than 1 species (Biham et al. 2001, Green et al. 2001,
Caselli et al. 1998). In the case of hydrogen, rate equations are
always valid because half of the chemisorbed sites are filled with
hydrogen atoms. The problem is different for deuterium since most of
the small grains are covered by less than 1 D atom.

\section{Validity check of the Rate Equation method}
\label{svalidity}
\begin{figure*}
\includegraphics[width=0.5\textwidth]{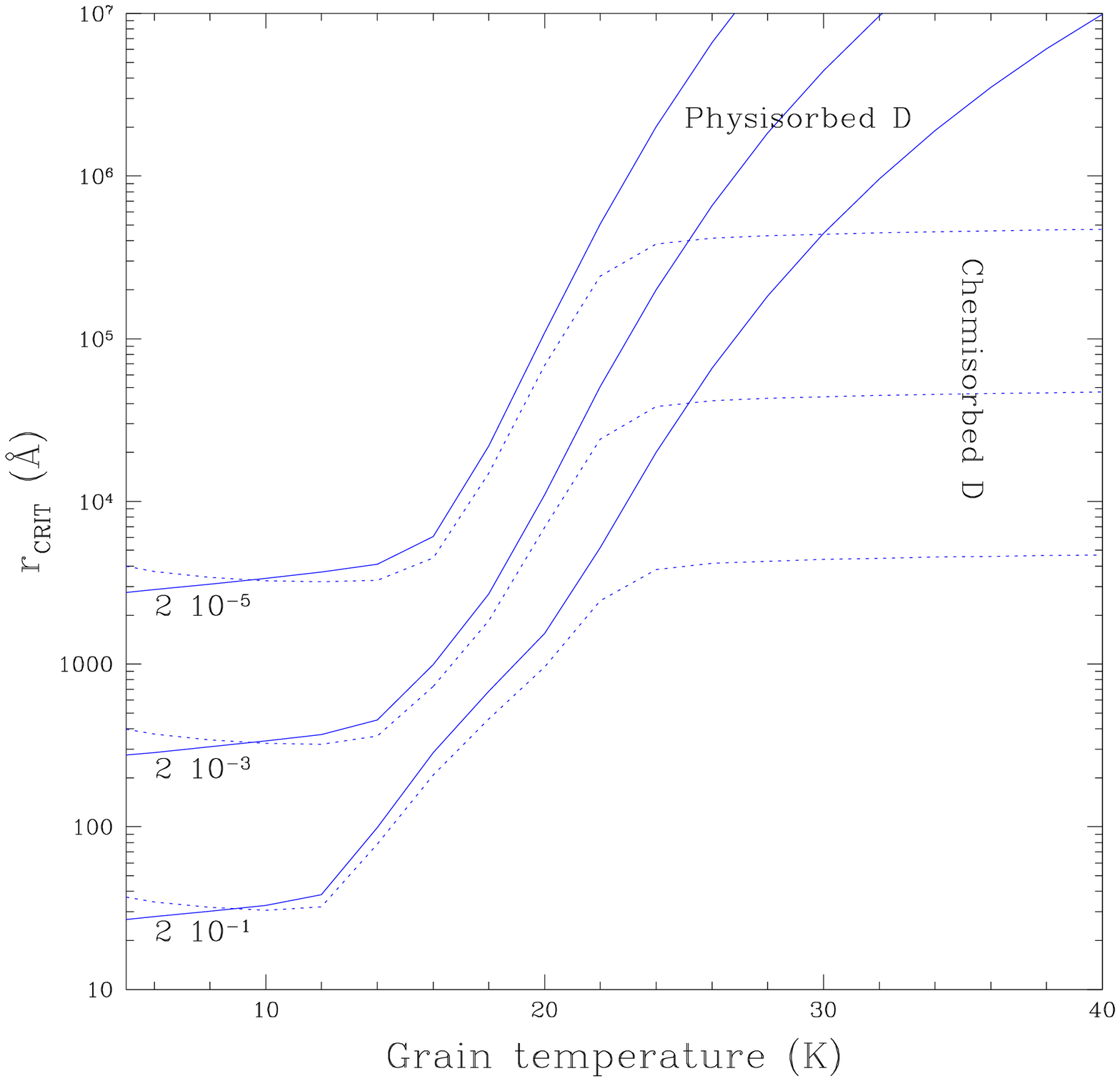}
\hspace{0.5cm}
\includegraphics[width=0.5\textwidth]{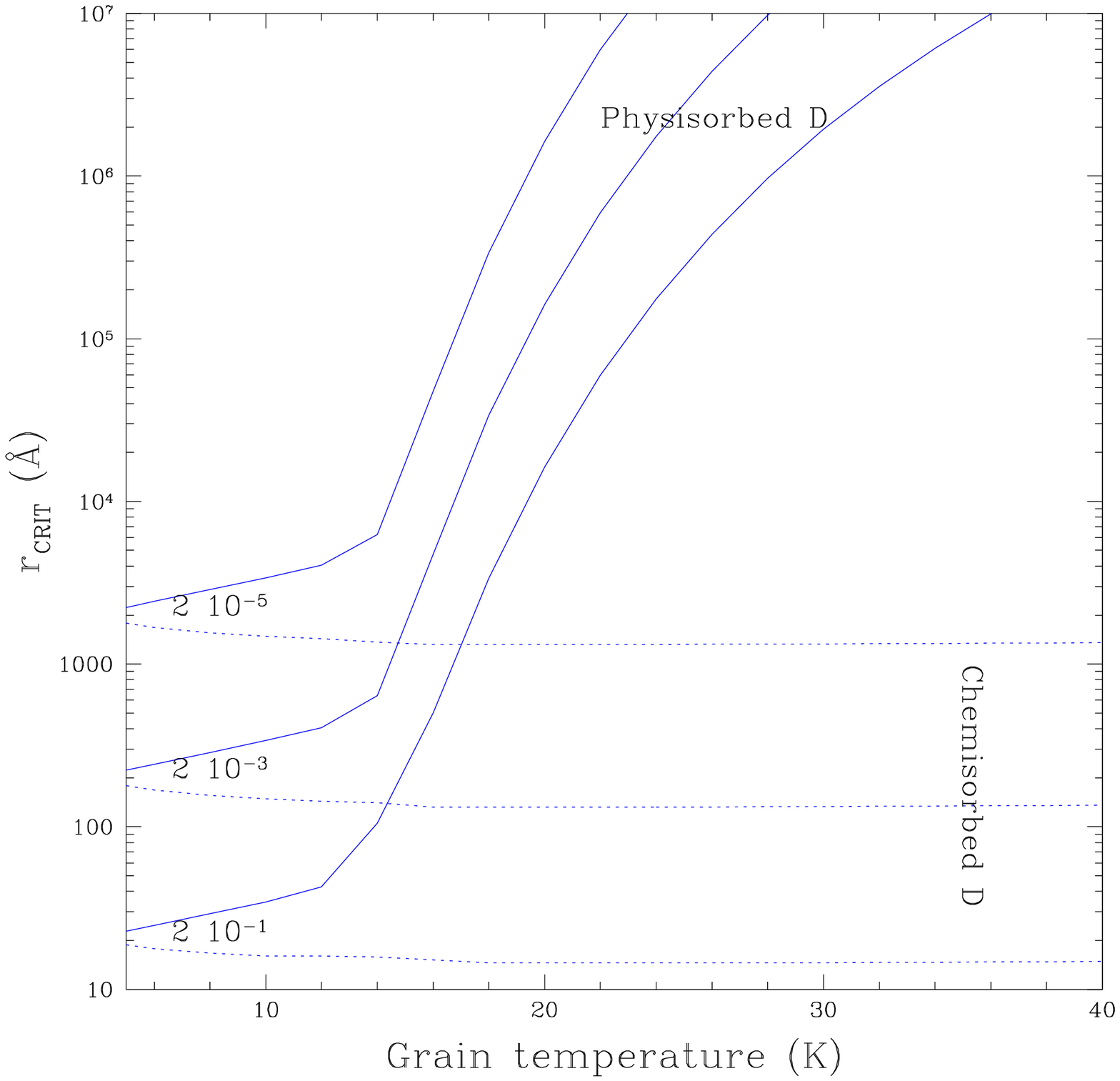}
\caption{Critical grain size for which a grain possess at least one
deuterium physisorbed and chemisorbed for graphitic (left panel) and
carbonaceous (right panel) grains.}
\label{crit}
\end{figure*}

For very small grains, while the number of hydrogen on the surface
(half of the chemisorbed sites) is always larger than one, and
therefore the rate equation method is always appropriate to describe
\hm\ formation, the number of D atoms on the grain can be less than
1. In figure ~\ref{crit}, we calculate the ``critical'' size of a
grain for which one deuterium is present on its surface. These
calculations have been performed for a density of nH=100 cm$^{-3}$, a
grain temperature of 15K, and for three different D/H ratio: 2$\times
10^{-5}$, 2$\times 10^{-3}$ and 2$\times 10^{-1}$ for graphitic grains
(left panel) and amorphous carbon grains (right panel).

\subsection{Monte Carlo and approximations}

For grains below the critical size, the rate equation method cannot be
used to follow the formation of HD and \dm\ on very small grains. We
developed a Monte Carlo model in order to follow the formation of HD
and \dm\ on these grains. In our model, the grain is seen as a squared
grid, with, at each intersection, the possibility to have a
chemisorbed and a physisorbed atom. We consider as before direct
chemisorption as well as Langmuir and Eley-Rideal mechanisms. When an
atom comes from the gas phase onto this grid, depending on its energy,
it can become physisorbed or chemisorbed. If the site is already
occupied, it can form a molecule which is released in the gas phase
(we consider $\mu$=0, which is a valid approximation for a range of
temperature at which newly formed molecules staying on the surface
thermally desorb). The position of the incoming atom on the grid is
chosen randomly. Once on the grid, the atom can move from site to site
according to its energy and follows a random walk. If two atoms arrive
in the same site, they associate to form a molecule which is released
in the gas phase. As discussed before, small grains ($\le$100\AA) are
considered to have surfaces similar to graphite, and big grains
similar to amorphous carbon.

\begin{figure*}
\includegraphics[width=0.5\textwidth]{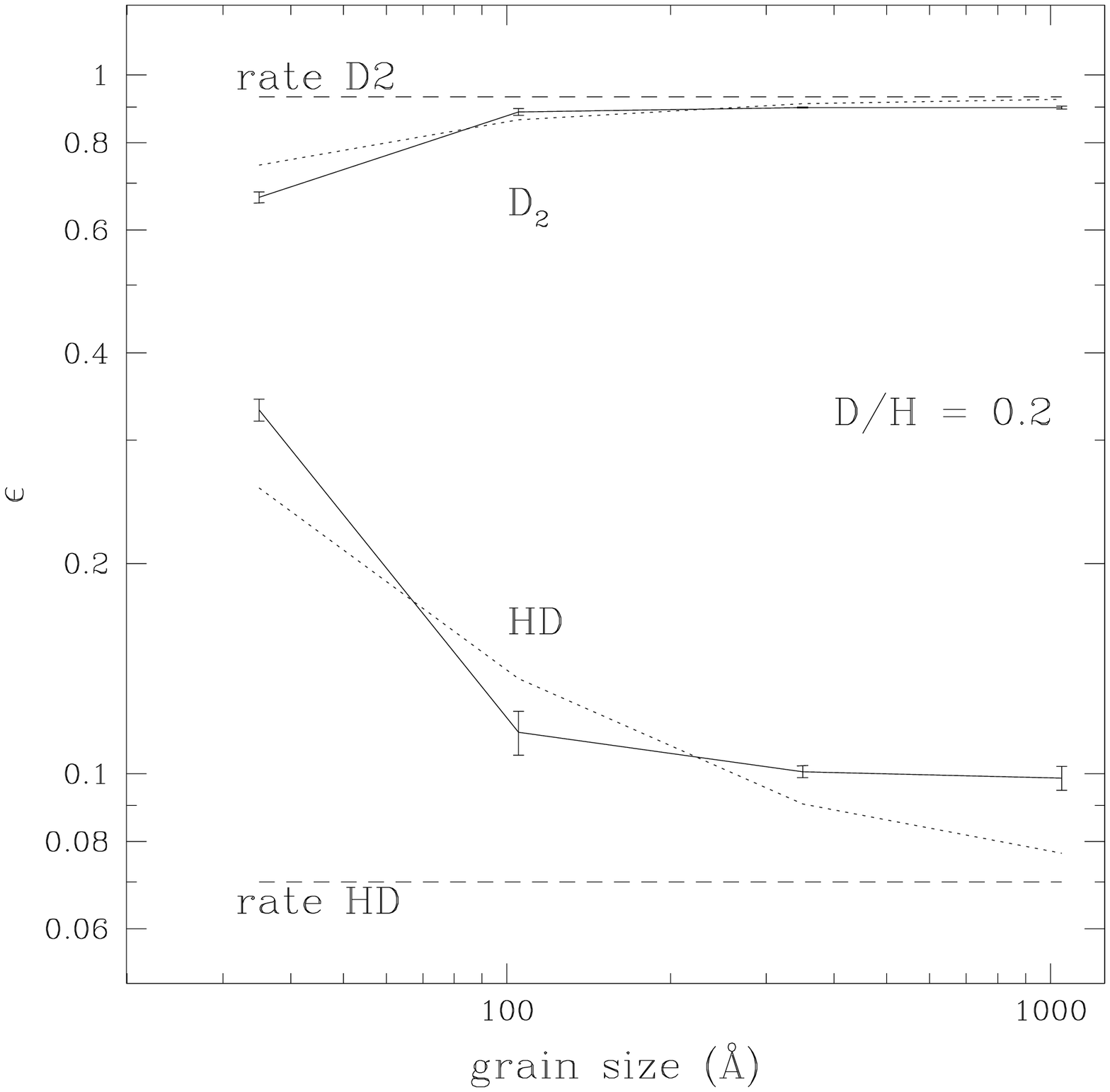}
\hspace{0.5cm}
\includegraphics[width=0.5\textwidth]{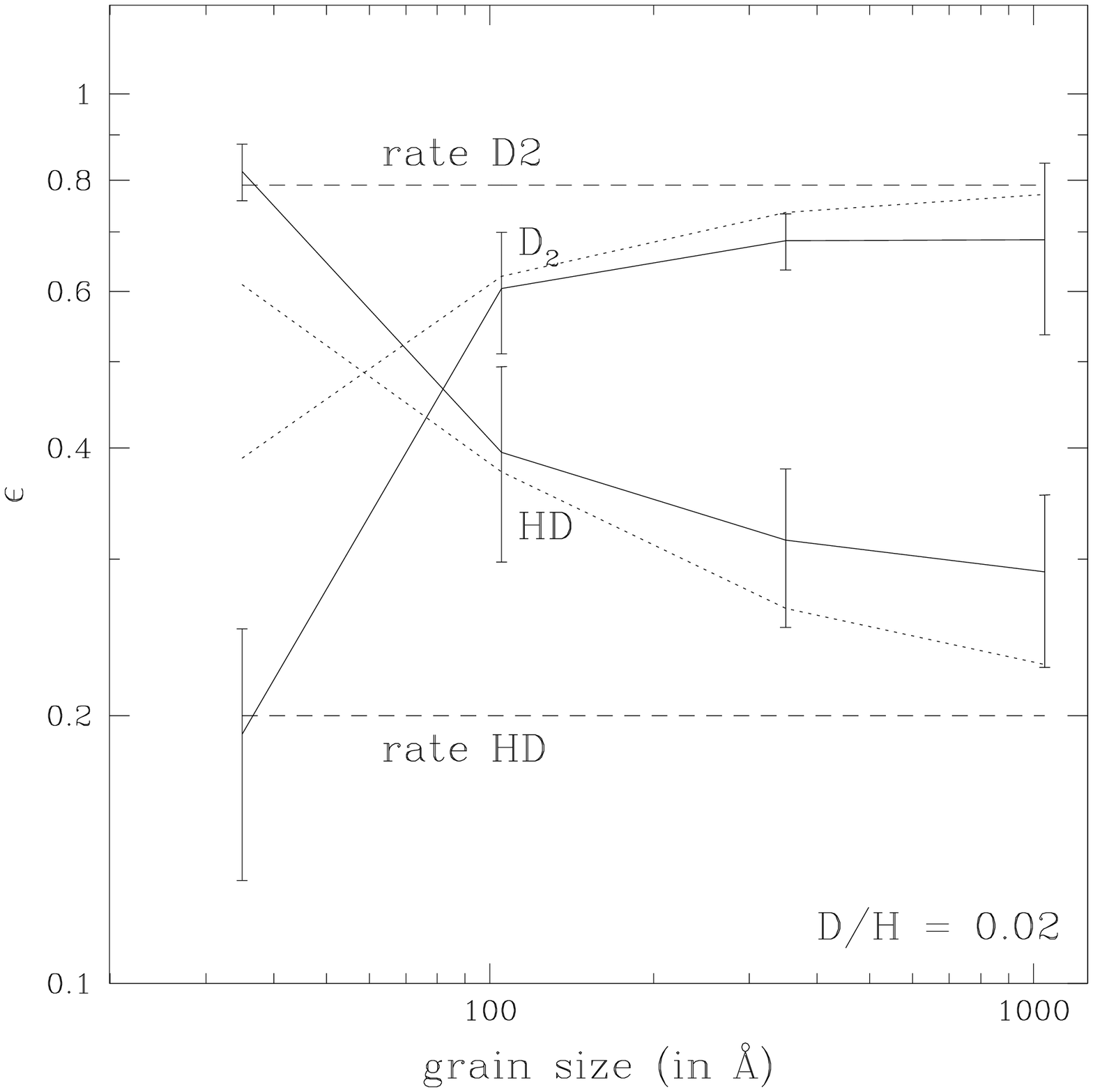}
\includegraphics[width=0.5\textwidth]{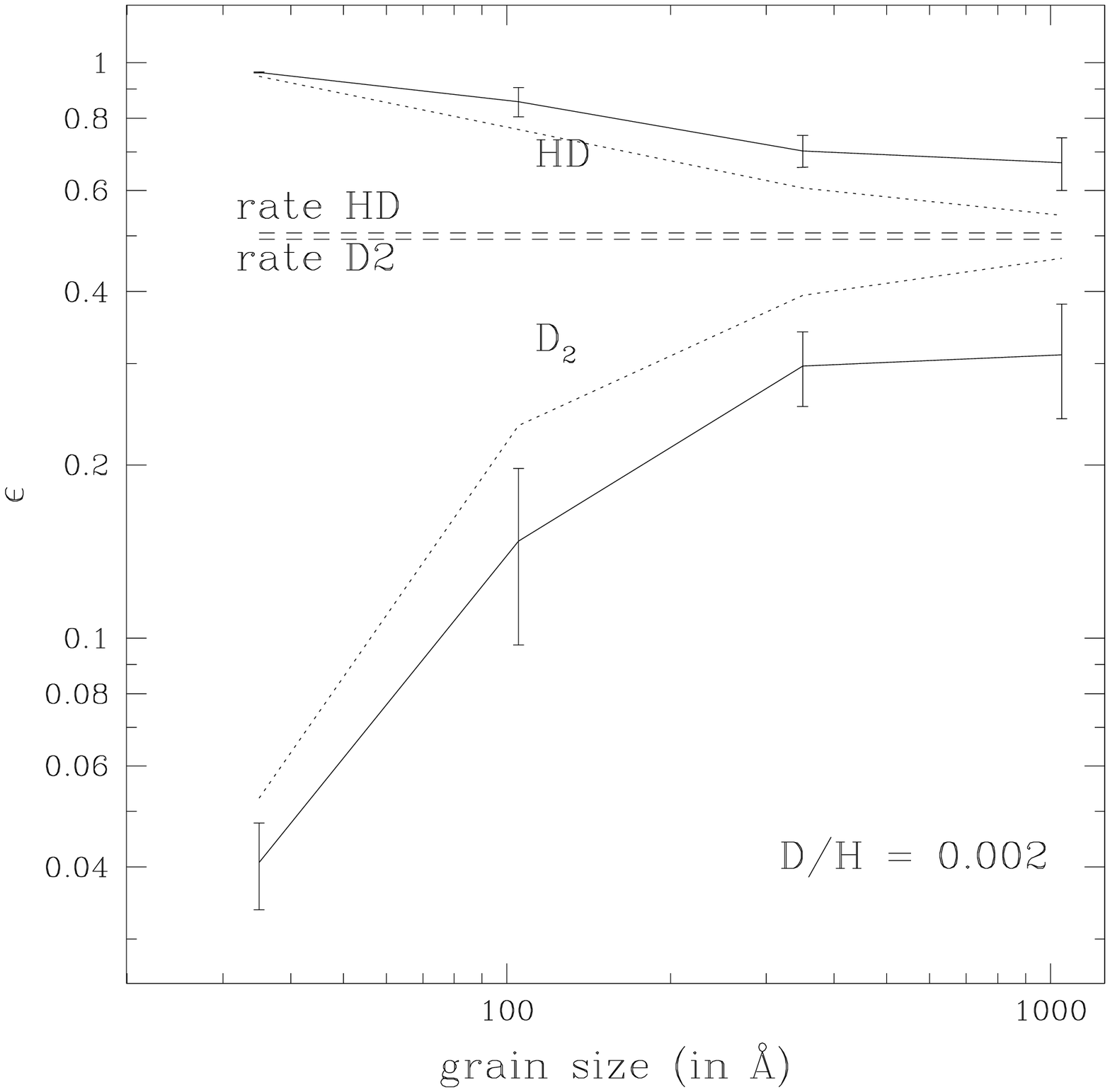}
\hspace{0.5cm}
\includegraphics[width=0.5\textwidth]{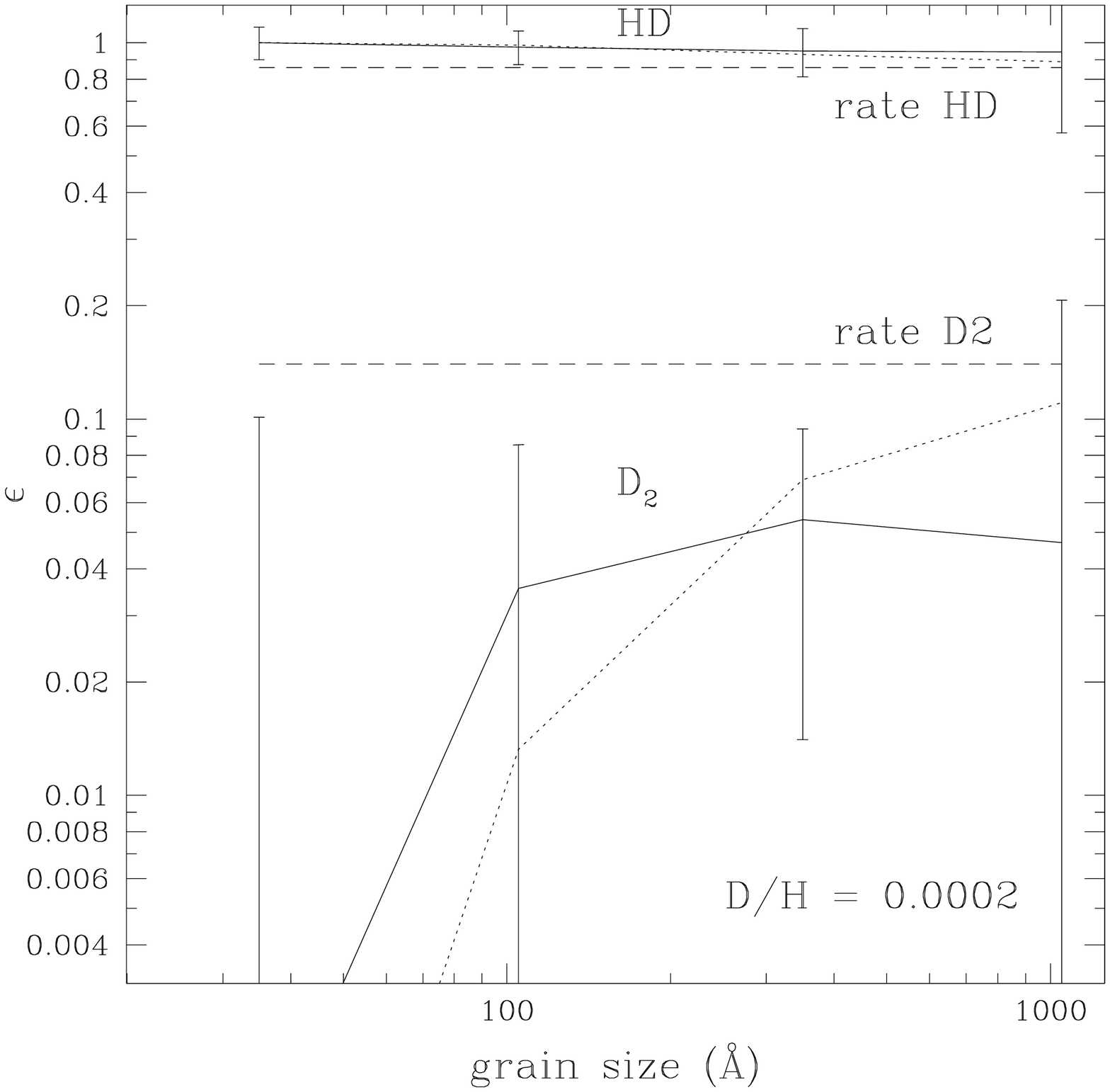}
\caption{Monte Carlo (solid lines) compared to rate equations (dashed
lines) simulations and our approximations (dotted lines) for the HD
and \dm\ formation efficiencies on graphitic surfaces. D/H ratio at
0.2 (top left), 2 10$^{-2}$ (top right), 2 10$^{-3}$ (bottom left) and
2 10$^{-4}$ (bottom right). The error bars represent the 95$\%$
confidence intervals.}
\label{mc1}
\end{figure*}

\begin{figure*}
\includegraphics[width=0.5\textwidth]{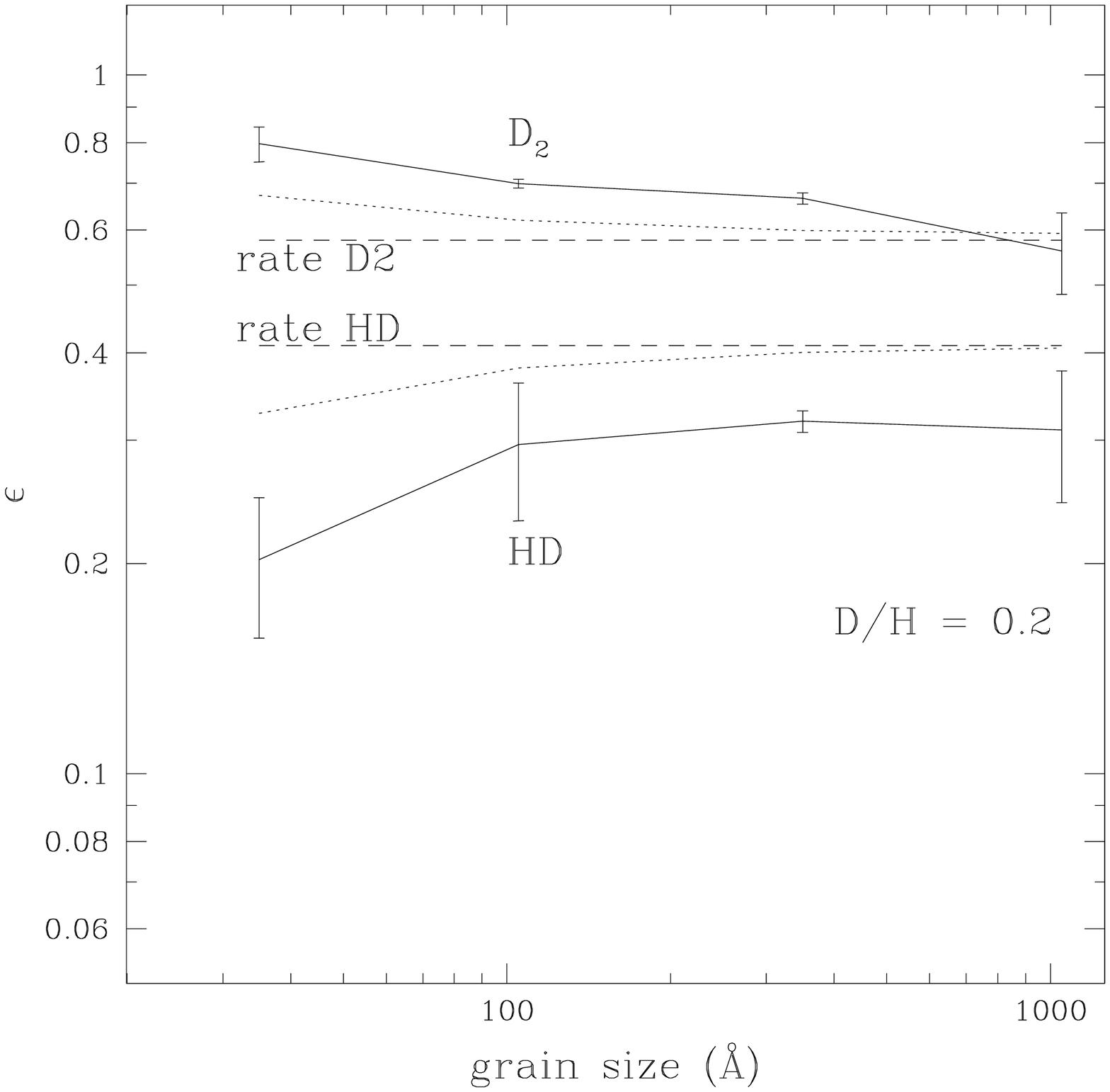}
\hspace{0.5cm}
\includegraphics[width=0.5\textwidth]{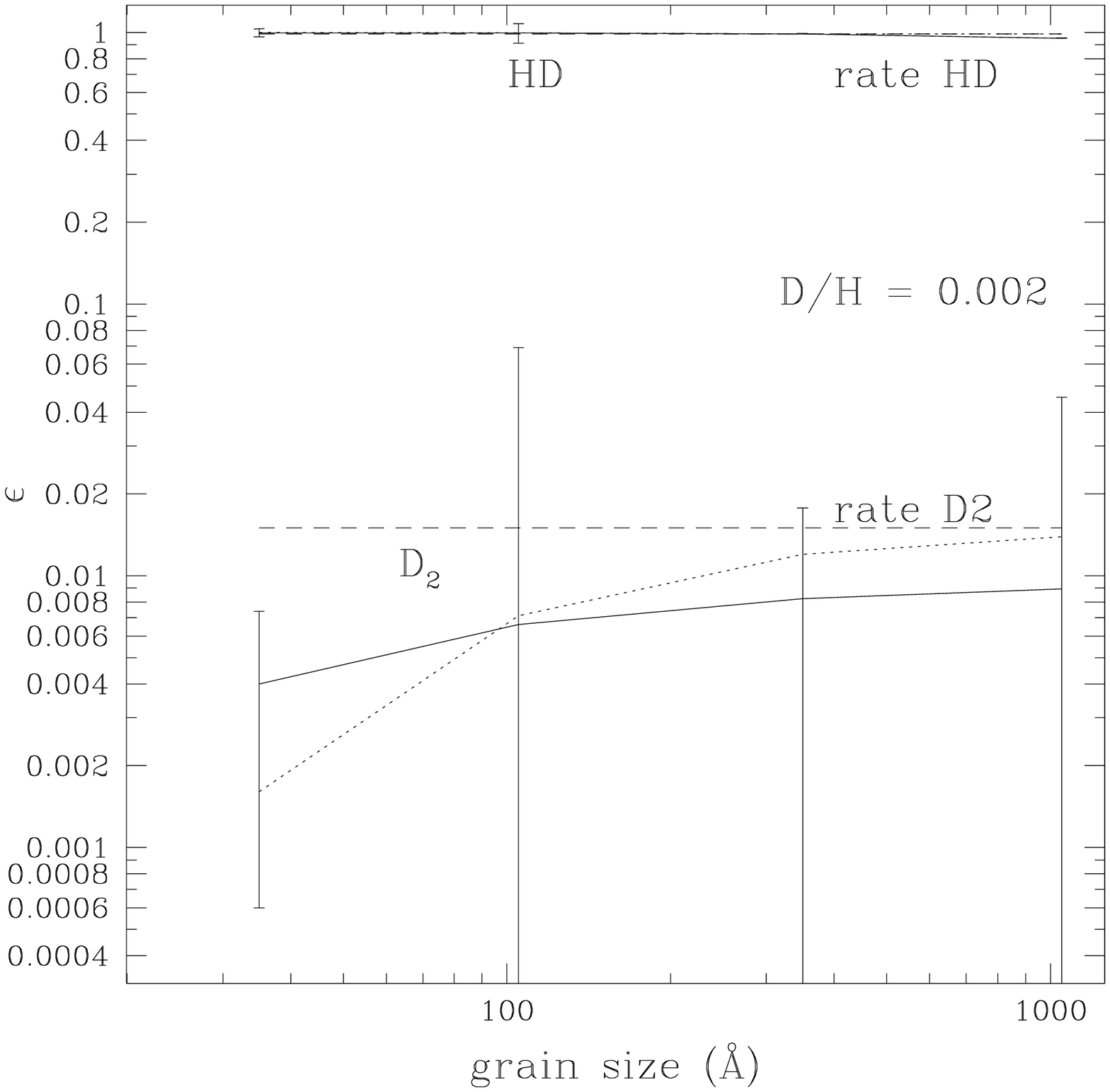}
\caption{Monte Carlo (solid lines) compared to rate equations
simulations (dashed lines) and our approximations (dotted lines) for
the HD and \dm\ formation efficiencies on amorphous carbon grains. The
D/H ratio is 2 10$^{-1}$ (left) and 2 10$^{-3}$ (right). The error
bars represent the 95$\%$ confidence intervals.}
\label{mc3}
\end{figure*}

\begin{figure*}
\includegraphics[width=0.5\textwidth]{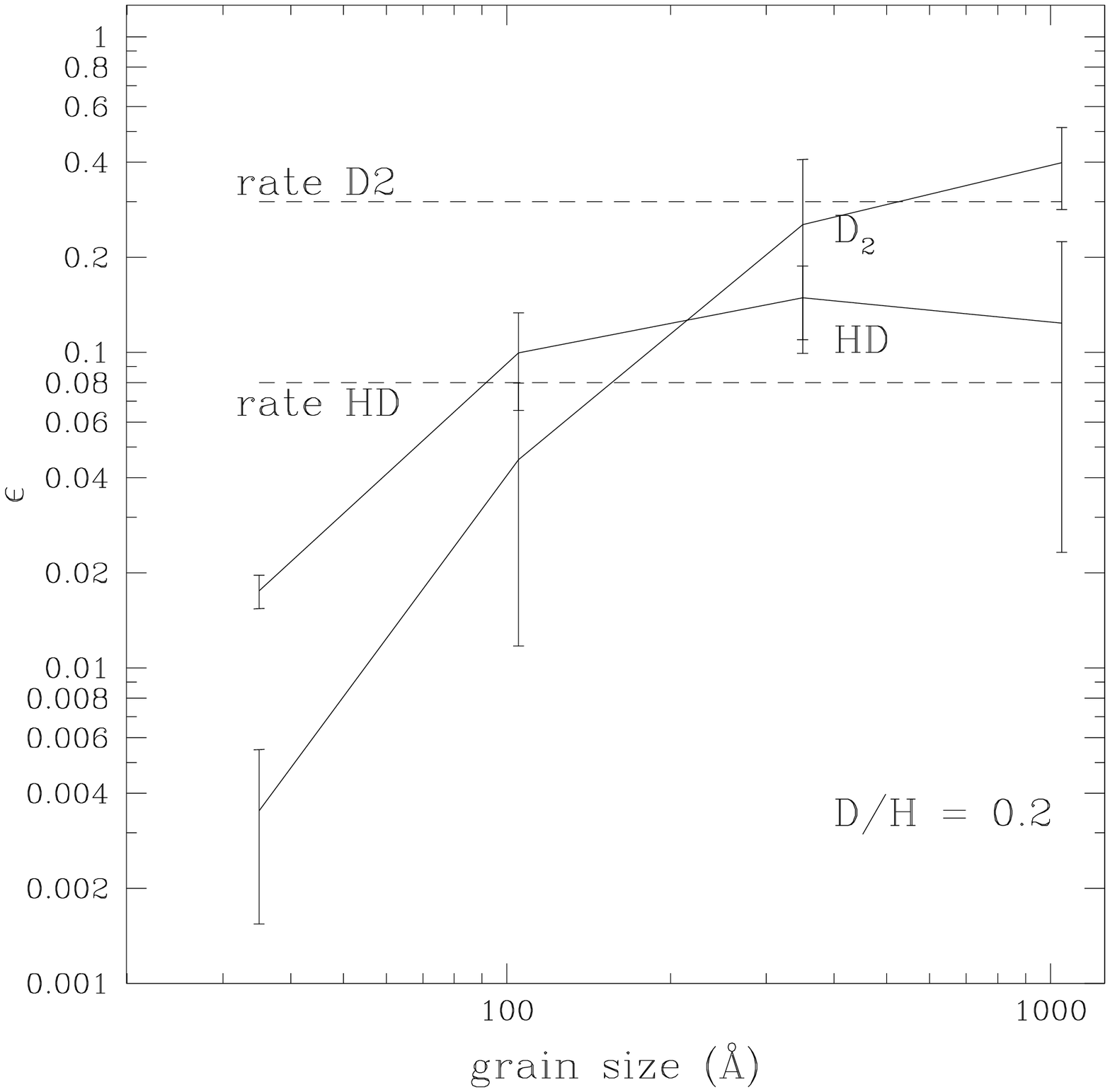}
\hspace{0.5cm}
\includegraphics[width=0.5\textwidth]{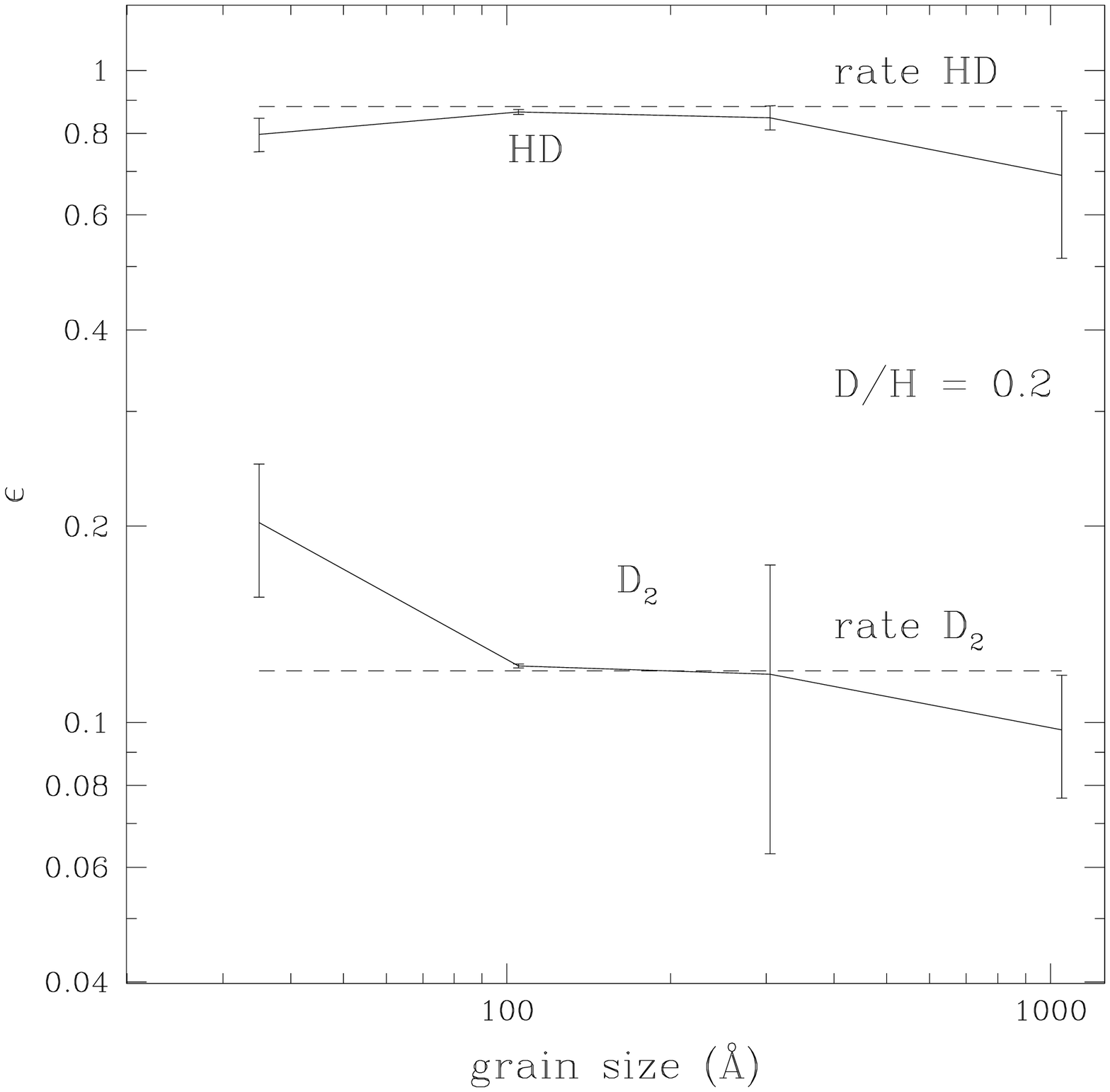}
\caption{Monte Carlo(solid lines) compared to rate equations
simulations (dashed lines) for the HD and \dm\ formation efficiencies
at higher grain temperatures. Left panel: graphitic surfaces at
20K. Right panel: carbonaceous surfaces at 25K.  The error bars
represent the 95$\%$ confidence intervals.}
\label{mc4}
\end{figure*}

Because the accretion of D on the grain is very low compared with the
accretion of H, the calculation times to form a HD and a \dm\ molecule
are very important. To deal with this problem, we performed the Monte
Carlo simulation for a D/H ratio from 10 to 1000 times higher than the
standard value 2$\times$10$^{-5}$, and extrapolate these results for
small D/H ratios. The calculations are performe here for
T$_{dust}$=15K, T$_{gas}$=100K and nH=100 cm$^{-3}$. Fig.~\ref{mc1}
show the HD and \dm\ efficiencies for graphitic grain from 100 sites
($\sim$ 30 \AA) to 10$^7$ sites ($\sim$ 1 $\mu$m) and for a D/H ratio
that varies from 2$\times$10$^{-4}$ to
2$\times$10$^{-1}$. Fig.~\ref{mc3} shows these efficiencies for
carbonaceous grains for a D/H ratio of 2$\times$10$^{-1}$ and
2$\times$10$^{-3}$. For big grains, steady state and Monte Carlo
simulations give the same results. As expected, when the number of
physisorbed deuterium on the grain is less than one, the two methods
diverge. This happens for a grain smaller than a critical size as
determined in Fig.~\ref{crit}. Below this critical size, with
decreasing grain sizes, the efficiency of \dm\ decreases, while the
efficiency of HD increases. Indeed, on big grains, the number of
physisorbed deuterium is higher or equal to 1, and a D atom coming
from the gas phase on the grain can meet another physisorbed
deuterium. On small grains, on the other hand, the number of
physisorbed deuterium can be less than 1, and a deuterium atom coming
from the gas phase on the grain can scout the surface before getting
trapped in a chemisorbed site without meeting another deuterium. When
getting chemisorbed, the deuterium can meet a H atom already present,
or can wait for a H atom coming in the chemisorbed site. According to
our calculations, as shown in fig~\ref{mc3} (right) and in
fig~\ref{mc4} (right), association of physisorbed atoms on grain
surfaces are dependent on the size of the grains, while association of
physisorbed and chemisorbed atoms are independent of size. {We
derived an empirical formula in order to reproduce the formation
efficiencies of HD and \dm\ with grain sizes}:

\begin{eqnarray}
\epsilon(D_2)&=&  \epsilon_{SS}(D_2)_{P+C}+ \epsilon_{SS}(D_2)_{P+P}\exp{\frac{-1}{\sqrt{D/H*nsites}}} 
\end{eqnarray}
\begin{eqnarray}
\epsilon(HD)&=& \epsilon_{SS}(HD) + \epsilon_{SS}(D_2) - \epsilon(D_2)\\ \nonumber
&=&  \epsilon_{SS}(HD) + \epsilon_{SS}(D_2)_{P+P} (1- \exp{\frac{-1}{\sqrt{D/H*nsites}}})
\end{eqnarray}
where $\epsilon_{SS}$ is the total formation efficiency in steady
state (see Fig.~\ref{dh}), $\epsilon_{SS}()_{P+C}$ is the formation
efficiency through the association of physisorbed and chemisorbed
atoms and $\epsilon_{SS}()_{P+P}$ the formation efficiency through the
association of two physisorbed atoms. Efficiencies due to the
formation of molecules through the association of 2 physisorbed atoms
decreases with decreasing grain size. textbf{These approximations are
compared to our Monte Carlo simulation in fig.~\ref{mc1} and
\ref{mc3}}.{ The error bars in our Monte Carlo simulations are
obtained by repeating the calculations at least twice, depending on
the CPU times required. The mean values are calculated together with
the standard deviation. This latter quantity is multiplied by the
appropriate student variable and the result represents the half
confidence interval at the critical risk of 5$\%$ (95$\%$ confidence
interval)}.

\subsection{Comparison with previous work}

 Recently, Lipshtat et al. (2004) studied the formation of HD and \dm\
 on grain surfaces. Their work differs from ours in several
 respects. First, their study considers only physisorbed sites and, as
 a consequence, they find \hm\ formation to be relatively inefficient
 at temperatures as low as 18 K (though note the efficiency is close
 to 1 below 16K). Second, they consider the formation of \hm, HD and
 \dm\ on grains as small as 50 \AA\ with a master equation approach,
 and integrate over a realistic grain size distribution. Their
 conclusions are that the production of the different molecules
 decreases with grain sizes. Also, when integrated over a large range
 of grain sizes, the ratio of the production of HD over that of \hm\
 \it{R}(HD)/\it{R}(\hm) \rm and \dm\ over \hm\ \it{R}(\dm)/\it{R}(\hm)
 \rm can be enhanced by a factor as high as 44 (78) for HD (\dm) (but
 sensitively dependent on T$_{dust}$). This enhancement is due to a
 decrease of production of \hm.

\begin{figure*}
\includegraphics[width=0.5\textwidth]{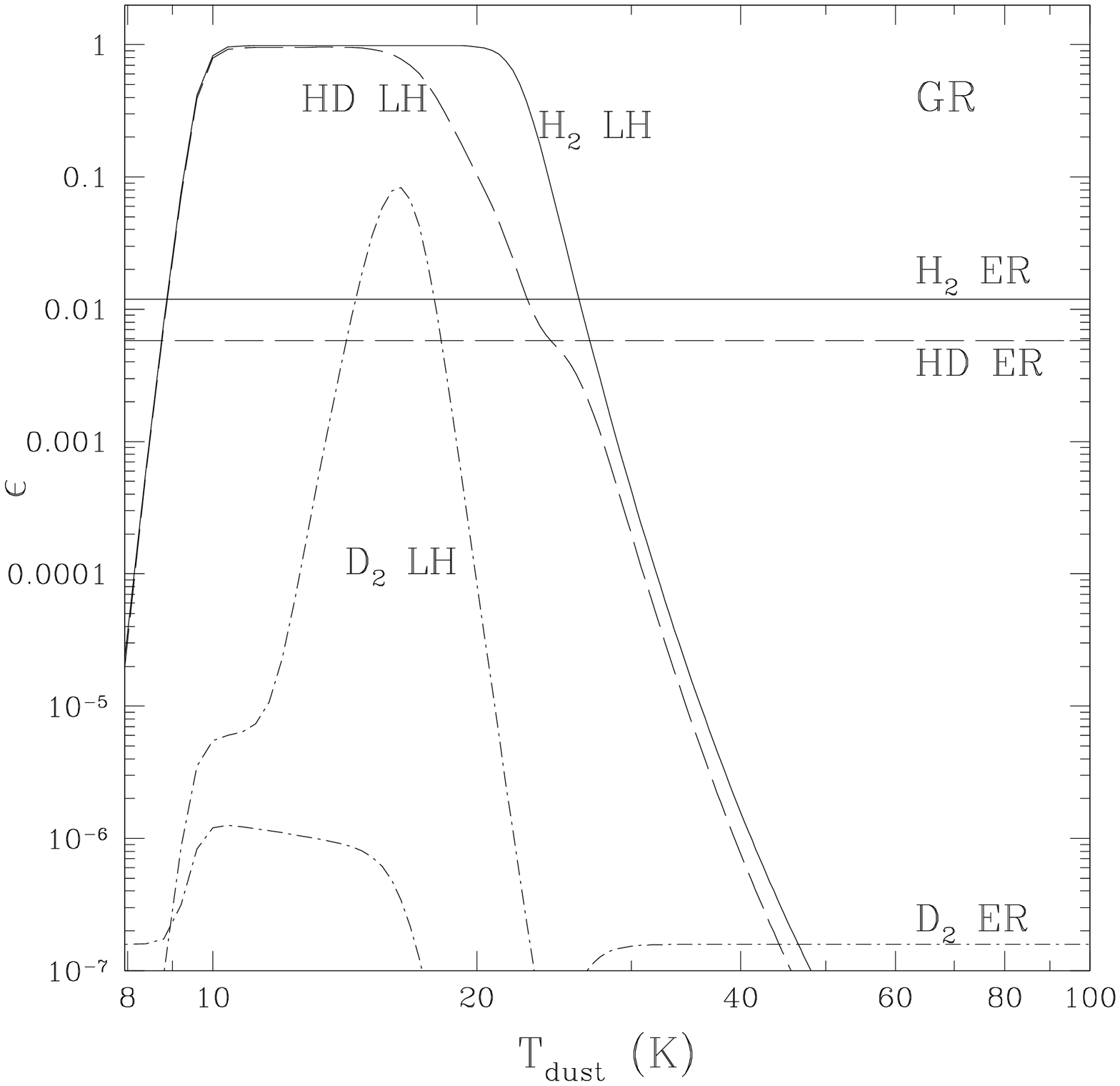}
\hspace{0.5cm}
\includegraphics[width=0.5\textwidth]{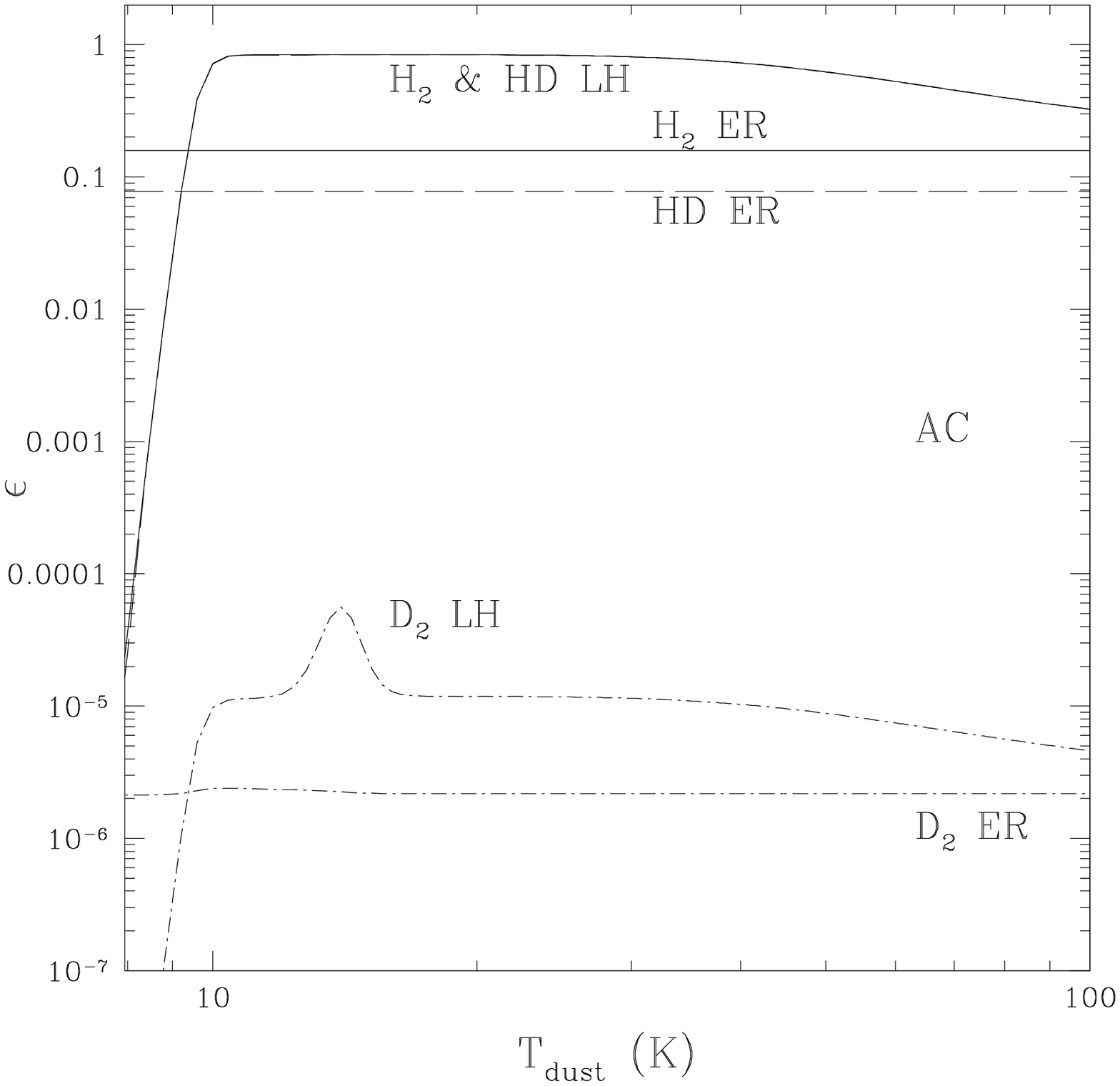}
\caption{\hm (solid lines), HD (dashed lines) and \dm\ (dotted dashed
lines) formation efficiencies on graphitic surfaces (left) and
amorphous carbon surfaces (right). LH Langmuir--Hinshelwood kinetic
and ER: Eley-Rideal mechanism.}
\label{Eff}
\end{figure*}

Our results differ in that, as a consequence of the presence of
 chemisorbed sites, we find almost 100 percent efficiency in both \hm\
 and HD formation (see Fig.~\ref{Eff}), as long as the formation is
 made through the association of physisorbed and chemisorbed
 atoms. With our model, \hm\ formation is independent of the size of
 the grains. \dm, which is mostly formed through the association of
 physisorbed atoms, has an efficiency which is grain size
 dependant. The efficiency of \dm\ formation from Figs.~\ref{Eff}
 and~\ref{thnh} can be enhanced by 3000 for graphitic grains and a
 factor of 100 for carbonaceous grains. Also, using a Monte Carlo
 simulation, we show that the formation of \dm\ decreases with
 decreasing grain sizes, while the formation of HD increases. The
 production of these molecules integrated over a range of grain sizes
 is discussed in the next section.

\section{Discussion}

{Density Functional Theory (DFT) calculations, showed that an H
atom get chemisorbed on graphitic surfaces atop a C atom which has to
move from the surface towards this adsorbate (Jeloaica \& Sidis 1999,
Sha \& Jackson 2002). This phenomenon is called puckering, and creates
a barrier against chemisorption of 0.2 eV. Recent studies from Rougeau
et al. (2006) and Hornekaer et al.(2006) showed that once an H atom is
chemisorbed on graphite, and therefore that a C atom has puckered out
of the plane, the next coming H atom can chemisorb in certain
neighbouring site (para site) without barrier. Another study by
Bachellerie et al. (2007) shows that once an H atom is chemisorbed in
a para site, a second H atom can also chemisorb in the same site
without barrier and form a molecule. These results are of main
importance for the formation of molecules when the chemisorbed atoms
are involved. Because rate equations do not take into account the
detailed structure of the surface, we could not take this phenomenon
into account. In this study, we developed a Monte Carlo simulation in
order to understand how the formation of \hm, HD and \dm\ changes with
the size of the grain. We compared these simulation to the rate
equations model, and therefore had to consider the same surface
without taking into account the properties of the para sites. In a
future work we will incorporate these properties in our Monte Carlo
simulations and see how the formation efficiencies of \hm\ and its
deuterated forms differ if we take into account the para sites. The
first effect of the inclusions of the para sites in our model should
be a more efficient \hm\ formation rate temperatures higher than 20K.}

\begin{figure*}
\includegraphics{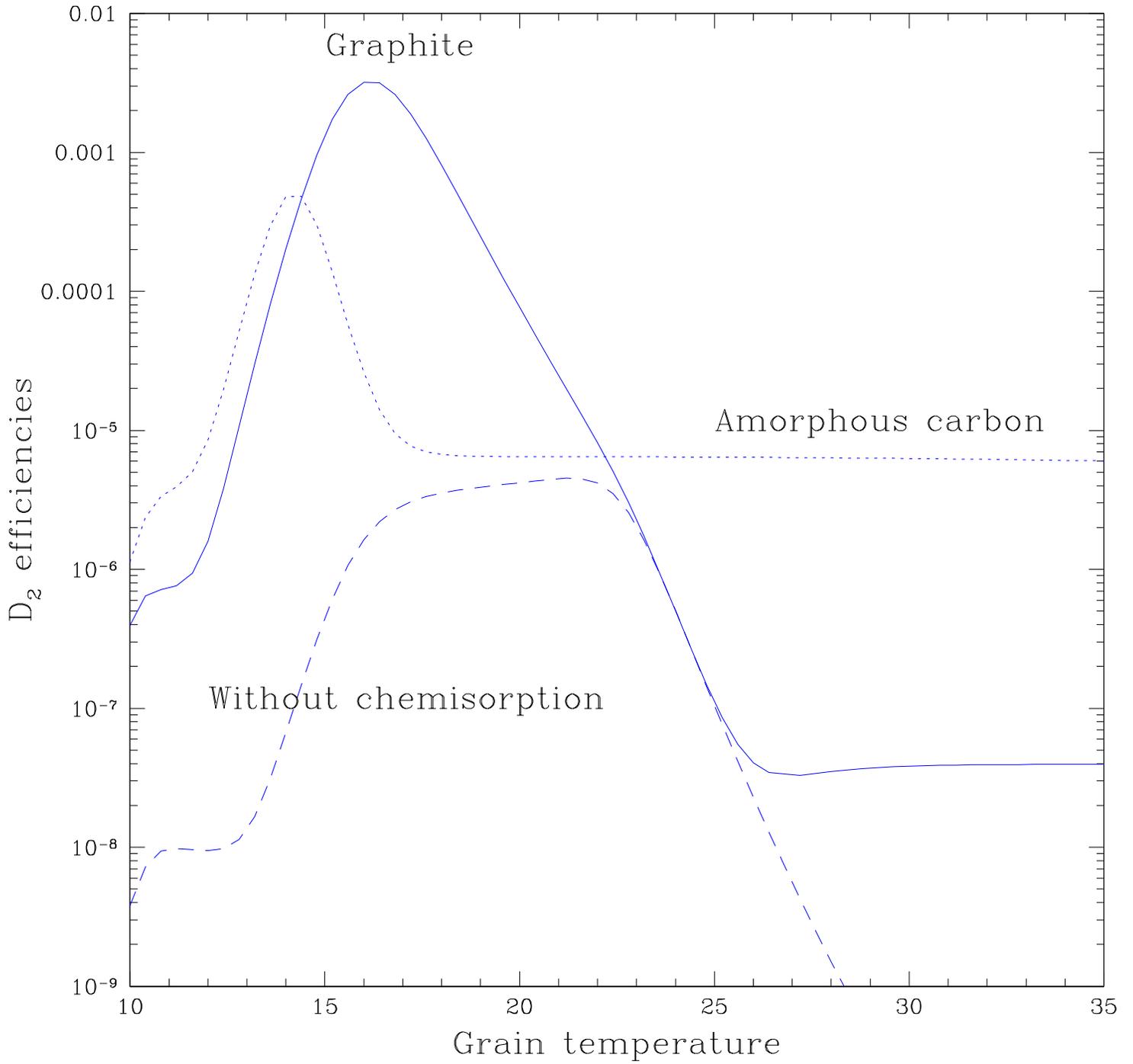}
\caption{Formation efficiency of D$_2$ molecules for graphite and
amorphous carbon and for surfaces without chemisorption sites, such as
icy grains.}
\label{d2eff}
\end{figure*}

To summarise the results obtained so far for \dm, figure~\ref{d2eff}
shows the variation of the D$_2$ formation efficiency with dust
temperature for the following parameters: $n(H_2 )$ = 1000 cm$^{-3}$
and D/H = 2 10$^{-5}$. The efficiency of D$_2$ formation varies
strongly from one kind of surfaces to another. If we consider surfaces
where only physisorption is possible, the formation of \dm\ is very
low, as discussed in Lipshtat et al. 2004. The inclusion of
chemisorbed sites in our model increases the formation efficiency of
\dm\ because of the different behaviour of H and D on grain
surfaces. H atoms are more mobile and will get easily trapped on
chemisorbed sites, whereas D will mostly stay in physisorbed
sites. Therefore, the surface structure has a big impact on the
formation of \dm. Surfaces, such as graphitic surfaces, present a very
high barrier against chemisorption. H atoms will overcome this barrier
by tunnelling, while D will tunnel much less efficiently, and stay in
the physisorbed sites. Since \dm\ forms mostly through the association
of physisorbed atoms, this segregation favours the formation of
\dm. Carbonaceous grains, on the other hand, present a very low
barrier against chemisorption, making the behaviour of H and D
similar. The formation efficiency of \dm\ present a small
enhancement. Therefore, deuteration will be more efficient on
graphitic surfaces (small grains) than on amorphous carbon surfaces
(big grains).

\begin{figure*}
\includegraphics{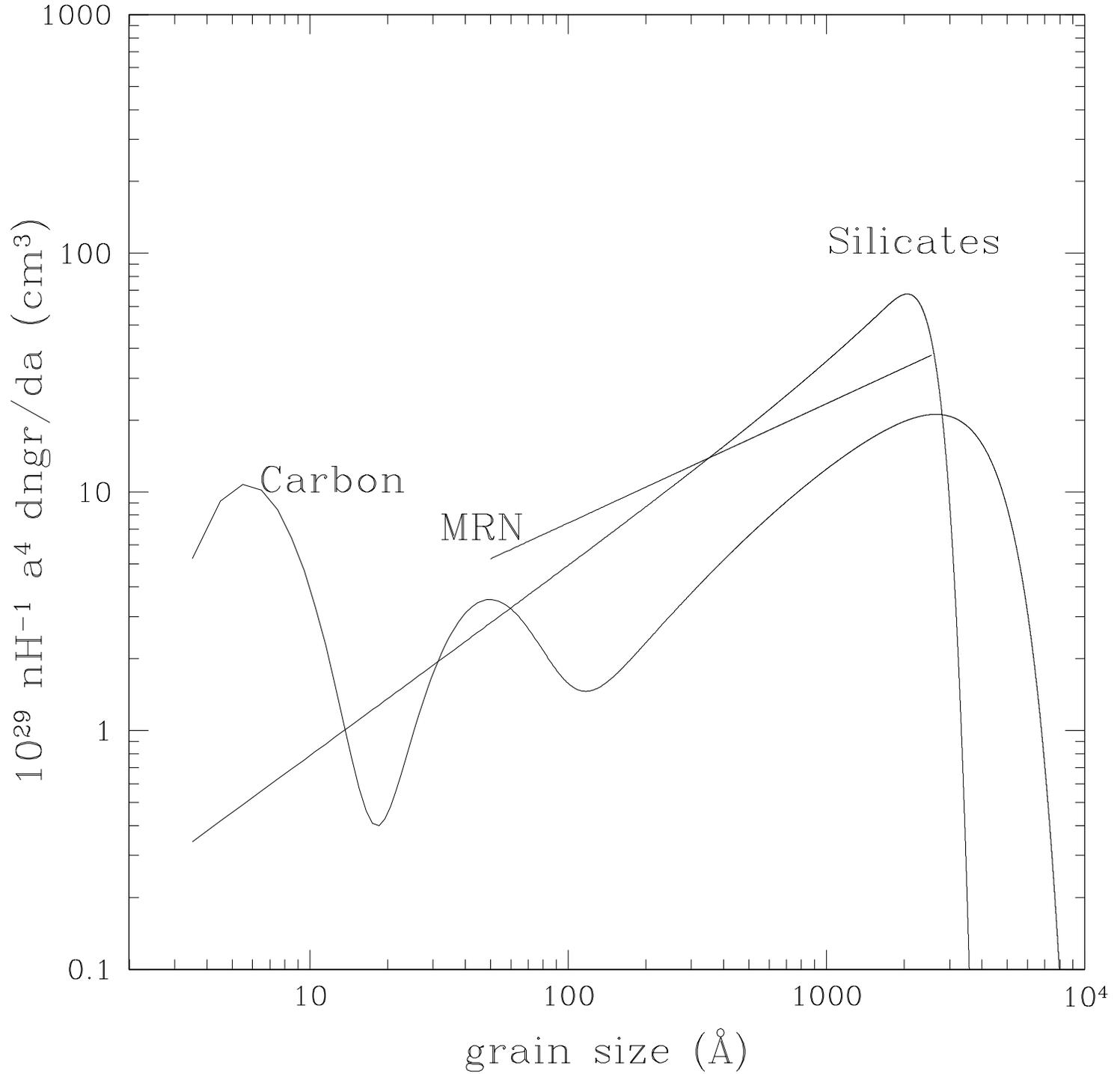}
\caption{Grain size distribution for carbon and silicate grains, as
estimated by Weintgartner \& Draine (2001).  Small grains
($\le$100\AA) are similar to PAHs, and big similar to amorphous carbon
grains.}
\label{dist}
\end{figure*}

\begin{figure*}
\includegraphics{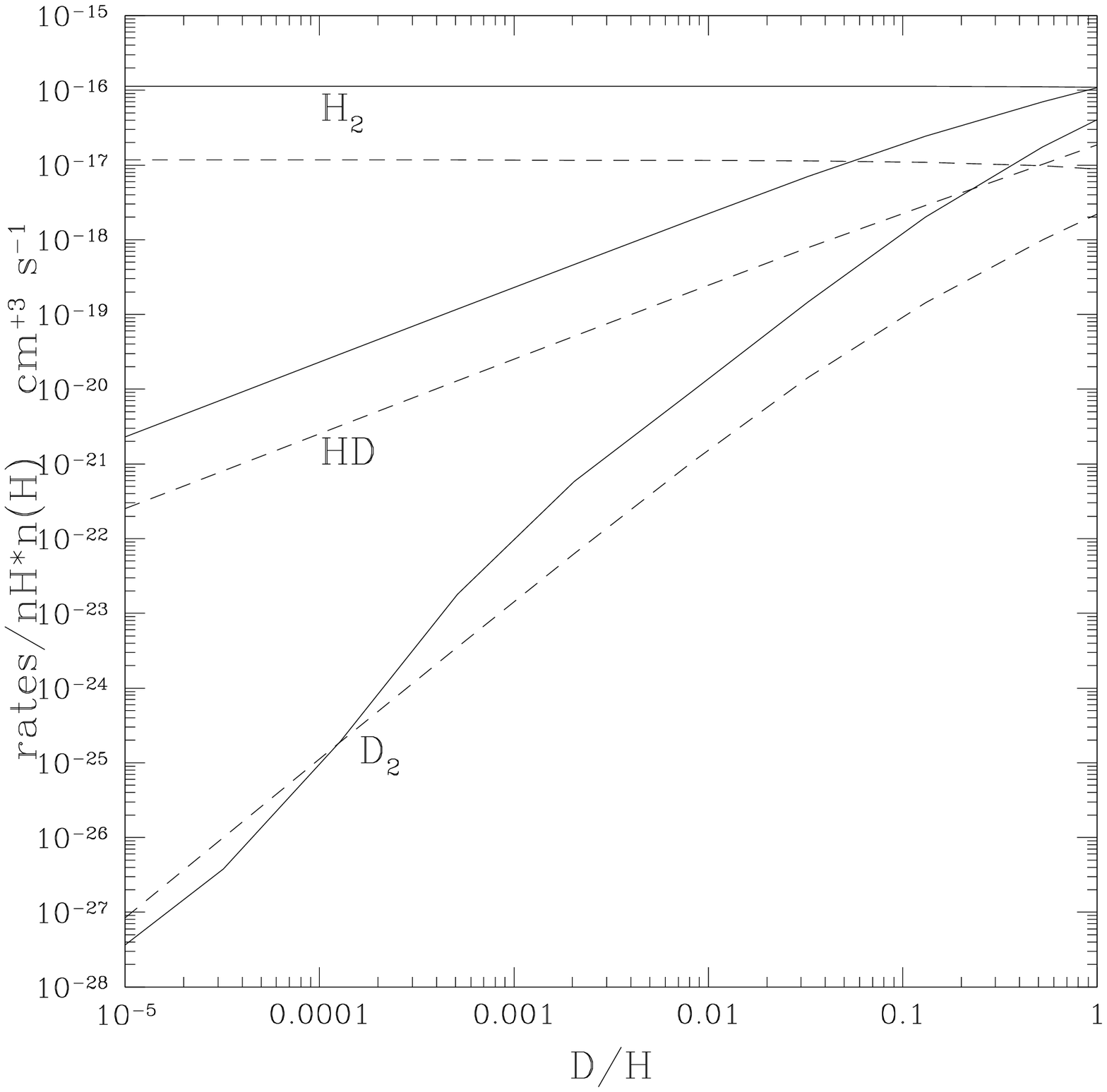}
\caption{Formation rate of \hm\ HD and \dm\ on small grains and PAHs
(solid lines) and big grains (dashed lines) as function of the D/H
ratio..}
\label{Rdh}
\end{figure*}

\begin{figure*}
\includegraphics{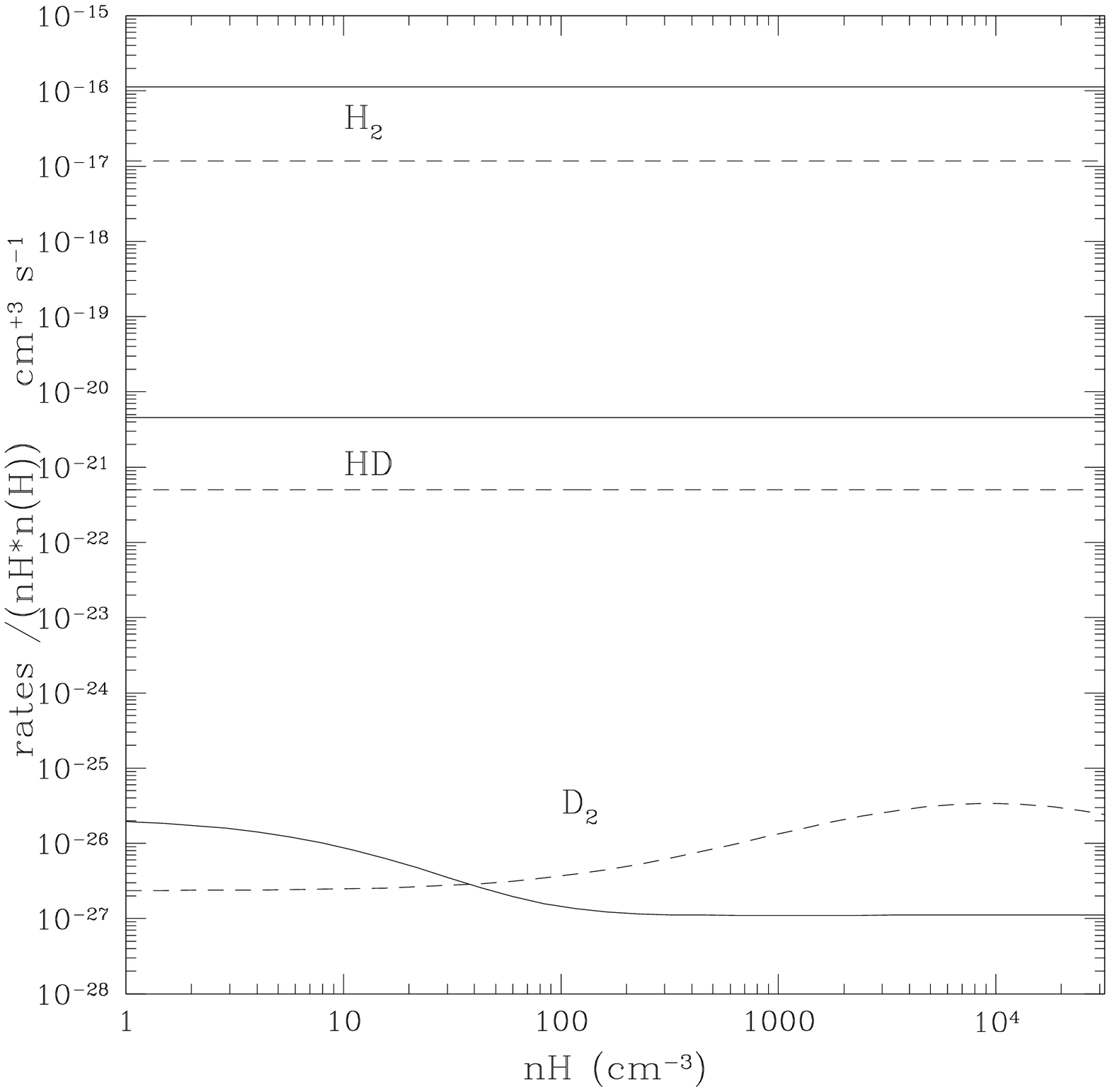}
\caption{Formation rate of \hm\ HD and \dm\ on small grains and PAHs
(solid lines) and big grains (dashed lines) as function of the
density.}
\label{Rnh}
\end{figure*}

\begin{figure*}
\includegraphics{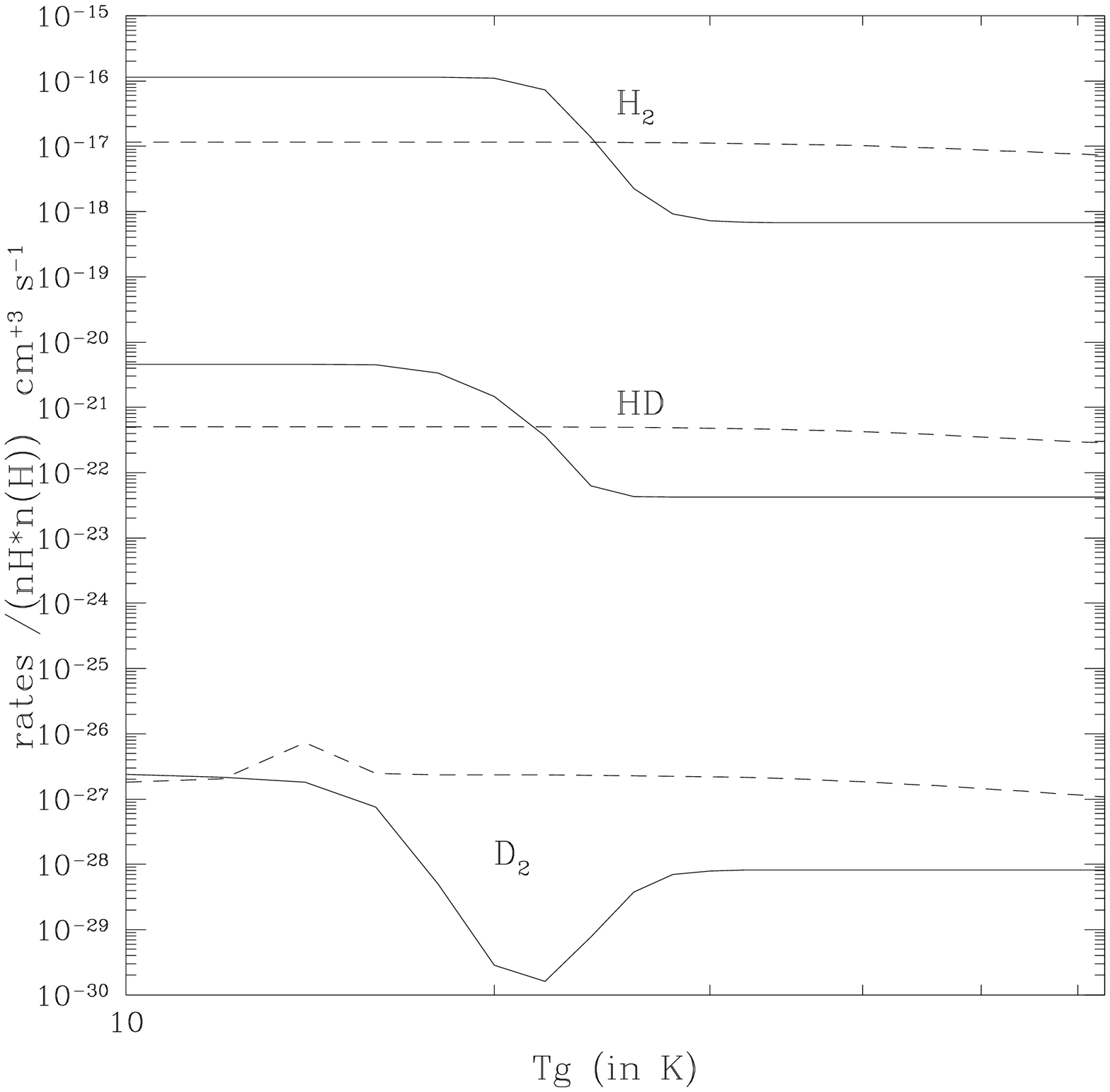}
\caption{Formation rate of \hm\ HD and \dm\ on small grains and PAHs
(solid lines) and big grains (dashed lines) as function of grain
temperatures.}
\label{Tg}
\end{figure*}

We consider a grain size distribution, as described by Weintgartner \&
Draine (2001), in order to predict the formation rate of \hm\, HD and
D$_2$ (in cm$^{-3}$ s$^{-1}$). This distribution is represented
figure~\ref{dist}. We calculated the total cross section of the
grains, for graphitic grains and carbonaceous grains. We considered
that PAHs and very small grains ($\le$ 100 \AA) have surfaces similar
to graphitic surfaces, and that bigger grains have surfaces similar to
carbonaceous surfaces. Then, we integrated the efficiencies for the
formation of the different molecules, with our correction for small
sizes, over the range of grain sizes, following the distribution of
Weintgartner \& Draine (2001). Figure~\ref{Rdh} shows how the
formation of these species depends on the D/H ratio for graphitic
surfaces (solid lines) and for carbonaceous grains (dashed lines), for
a grain temperature of 15K, a gas temperature of 100 K and a density
of 100 cm$^{-3}$. The formation of \hm\ and HD under these
circumstances comes mostly from the contribution of small grains (Gr
grains). On the other hand, \dm\ is mostly formed on big grains (AC
grains), for low D/H ratios (D/H $\le$ 10$^{-4}$), and on small grains
(Gr grains) for higher D/H ratio. Figure~\ref{Rnh} presents the same
results as a function of the density, for a grain temperature fixed at
15K and a gas temperature of 100K. While the formation of \hm\ and HD
is mostly due to the small grains contribution for any n(H), \dm\ is
mainly produced on small grains at low densities and on big grains at
high densities. Figure ~\ref{Tg} present the formation rate of the
different species as a function of the grain temperature, for a gas
temperature of 100 K, a density of 100 cm$^{-3}$ and a D/H ratio of
2$\times$10$^{-5}$. First, it seems clear that the formation of \hm\
comes from the contribution of small grains at low grain temperature,
and of big grains at higher grain temperature. These two
contributions, added together, reproduce quite well the formation
rates observed in different PDRS, as described by Habart et
al. (2003). Small grains contribute strongly on the formation of \hm\
and HD at low grain temperatures. Then, for temperatures higher than
$\sim$ 20K, these molecules are formed mostly on big grains. At these
higher temperatures, the rate of \hm\ formation drops to 10$^{-17}$ nH
n(H) cm$^{-3}$ s$^{-1}$. This value is similar to the one observed in
PDRs where warm dust grains are present as in NGC 2023 and the Orion
Bar (Habart et al. 2004, Allers et al. 2005). For the case of \dm, for
standard D/H, its formation comes from the big grains.

{In the interstellar medium, dust is heated by photons. Dust
grains receive a short heat impulse, resulting in temperature
fluctuations, especially for small grains. This phenomenon has been
described by Draine \& Li (2001) who showed that grains of the size of
50 \AA\ can fluctuate from 5 to 40 K, but still spend most of their
life (90$\%$) at temperatures below 20K. For smaller grains (25\AA),
the fluctuations are even more important (from few K to 50 K), but the
grain spend 95 $\%$ of its life at temperatures below 20K. In this
study we did not consider the temperature fluctuations of small
grains. The limiting factor in the formation of molecules on very
small grains is the accretion. In environments with a density of 100
cm$^{-3}$, a small grain of 10 \AA\ (30 \AA) will receive a H atom
from the gas phase every 7 10$^6$ (7 10$^5$) seconds and a D atom
every 3 10$^{11}$ (3 10$^{10}$) seconds. Once the atom arrives on the
grain, it can evaporate (if T$_{dust} \ge$ 20K for graphitic surface,
and for very high temperatures for amorphous carbon surface, see
fig.\ref{trans}), or become chemisorbed in a few hundreds of seconds
(on graphitic surfaces) and a few tens of seconds (on amorphous carbon
surfaces). This time can be estimated by $\frac{1}{alpha_{pc}}$, and
the number of physisorbed sites that the atom will visit can be
estimated by the ratio $\frac{\alpha_{pp}}{alpha_{pc}}$. Therefore, in
most cases an atom coming on a grain will become chemisorbed in an
empty or in a filled chemisorbed site. In the latter case, a molecule
will be formed. If the grain is at temperatures higher than 20K, then
the atom will evaporate and no molecule will be formed. Therefore, the
temperatures fluctuations of dust grains should reduce the efficiency
of the formation of the molecules by a maximum of 10$\%$ for grains
lower than 100\AA (which spend 90$\%$ of their life at temperatures
below 20K). A recent study by Cuppen, Morata \& Herbst (2006) shows
that when considering grains that possess only physisorbed sites, the
efficiency depends both on the modal temperature (the most frequent
temperature of the grain) and its fluctuations. In this case, these
fluctuations shuts down the formation of \hm\ on grains smaller than
100 \AA\ when rough surfaces are considered. In our case, because we
consider chemisorbed atoms, the formation of \hm\ will just slightly
decrease by a maximum 10$\%$ (for graphitic surfaces) or will not
decrease at all (for amorphous carbon surface). }

{Another important process that we did not take into account in
this study is the so-called ``Hot Atom Mechanism''. An atom that
becomes physisorbed on a dust grain is not directly in thermal
equilibrium with the grain. In most cases, the atom comes from the gas
phase with a higher energy, and once on the grain surface, it bounces
against potential walls and looses slowly its energy. Such a process
has been described by Buch \& Zhang (1991), and shows how many sites
an atom can scout before being in thermal equilibrium with the
grain. In our study, the efficiencies of the formation of the
different molecules is lower at low grain temperatures. Indeed, at
grain temperatures less than 10K, atoms visit grain surfaces through
tunnelling and when they encounter to form molecules, a fraction of
the newly formed molecules stay on the grain. Therefore, the reason of
a low formation efficiency is that dust grains are saturated with
molecules. In this study, the Hot Atom Mechanism will not increase the
efficiency of the formation of molecules at low grain temperature.}

\section{Conclusions}

We developed a rate equation model for the formation of \hm\ and its
deuterated forms on carbonaceous surfaces. This model takes into
account the structure of carbonaceous surfaces in order to describe
the formation mechanism of the different molecules. Small carbonaceous
grains ($\le$ 100\AA), also called PAHs, possess surfaces that are
similar to graphite, while big grains have surfaces similar to
amorphous carbon. One or the other type of grain, and therefore of
the surface, will have a big impact on the chemistry and on the
deuteration of \hm.

Small grains present surfaces characteristics similar to graphitic
surfaces which show a high barrier against chemisorption. Big grains,
on the other hand, have surfaces similar to amorphous carbon that
present no barrier against chemisorption. Because of their mass
differences, H and D atoms will overcome high barriers against
chemisorption with different efficiencies. H atoms can tunnel through
the barrier to populate the chemisorbed sites, while D atoms, with
higher mass, tunnel much less efficiently and therefore mostly
populate physisorbed sites. This segregation leaves the D atoms free
to travel and associate on physisorbed sites, while the H atoms are
trapped in chemisorbed sites. In this sense, small grains (graphitic
surface) favour deuteration.

The rate equation method is applicable only when there is at least one
species on the grain. For molecular hydrogen, this method is always
valid since a \hm\ molecule is formed through the association of a
physisorbed and a chemisorbed H atoms. Because half of the chemisorbed
sites are filled with hydrogen, there is always more than one H on the
surface, and therefore rate equations are always valid.

In the case of deuterium, because of the low D/H ratio in the ISM, it
is common that a grain possesses less than 1 D atom on its surface. To
understand the formation of HD and \dm\ in this case, we developed a
Monte Carlo simulation. Our results show that HD formation efficiency
increases with smaller grain sizes, while \dm\ decreases. Indeed, with
decreasing grain sizes, the number of physisorbed D decreases, and
therefore a D atom on a grain will easily get trapped in a chemisorbed
site, and form HD, instead of finding another physisorbed D, and form
\dm. We propose an approximation to describe the formation of HD and
\dm\ as a function of grain sizes.

We calculated the formation rate of \hm\ and its deuterated forms when
the grain sizes follow the Weintgartner \& Draine (2001)
distribution. We differentiated the contribution of small ($\le$
100\AA; Gr Grains) and big grains (AC grains) in the formation of the
different species. Our results show that \hm\ and HD are formed at low
grain temperature ($\le$ 25K) mostly on small grains (Gr grains), and
on big grains at higher grain temperatures (AC grains). The formation
of \dm, on the other hand, can be dominated by the contribution of big
grains, for high densities, and of small grains for low densities.

At last, this paper shed some light on the chemistry of species on
realistic dust grains, that possess physisorbed as well as chemisorbed
sites. On big grains certain molecules form through the association of
2 physisorbed atoms. On small grains, the physisorbed atoms are much
less abundant and become chemisorbed before finding another
physisorbed atom on the surface. Therefore, small grain will favour
the formation of molecules through the association of physisorbed and
chemisorbed atoms.

\acknowledgements{We would like to thank the referee for his
constructive comments that helped significantly in improving this
manuscript.}

\newpage

\end{document}